\documentclass[journal]{IEEEtran}
\usepackage{setspace}
\usepackage{enumitem}
\usepackage[dvipsnames]{xcolor}
\usepackage{graphicx}
\usepackage{epstopdf}
\usepackage{csquotes}
\usepackage{subfig}
\usepackage{mathtools}
\usepackage{amssymb, amsmath,amsthm, amsfonts}
\usepackage{ifthen}
\usepackage{tikz}
\usepackage{cite}
\usepackage{verbatim}
\usepackage{pifont}
\usepackage{bm}  
\usepackage{stackengine}
\usepackage{bbm}

\allowdisplaybreaks
\usepackage{accents}

\usepackage{xpatch}
\makeatletter
\xpatchcmd{\@thm}{\thm@headpunct{.}}{\thm@headpunct{}}{}{}
\makeatother

\DeclareMathOperator*{\argmax}{arg\,max}

\DeclareSymbolFont{bbold}{U}{bbold}{m}{n}
\DeclareSymbolFontAlphabet{\mathbbold}{bbold}

\newtheorem{theorem}{Theorem}
\newtheorem{cor}{Corollary}
\newtheorem{lemma}{Lemma}
\newtheorem{prop}{Proposition}

\newtheorem{remark}{Remark}

\newtheorem{definition}{Definition}

\newcommand\Item[1][]{%
  \ifx\relax#1\relax  \item \else \item[#1] \fi
  \abovedisplayskip=0pt\abovedisplayshortskip=0pt~\vspace*{-\baselineskip}}
\newcommand{\kl}[2]{\mathsf{D}_\mathsf{KL}\left(#1\middle\|#2\right)}
\newcommand{\tv}[2]{\delta_\mathsf{TV}\left(#1,#2\right)}
\newcommand{\typ}[3]{\mathcal{T}_{#2}^{(#3)}(#1)}

\newcommand{\cB}{\mathcal{B}}

\newcommand{\cH}{\mathcal{H}}
\newcommand{\cI}{\mathcal{I}}
\newcommand{\cJ}{\mathcal{J}}

\newcommand{\cM}{\mathcal{M}}

\newcommand{\cP}{\mathcal{P}}

\newcommand{\cU}{\mathcal{U}}
\newcommand{\cV}{\mathcal{V}}

\newcommand{\cX}{\mathcal{X}}
\newcommand{\cY}{\mathcal{Y}}
\newcommand{\cZ}{\mathcal{Z}}

\newcommand{\PP}{\mathbb{P}}

\newcommand{\RR}{\mathbb{R}}

\newcommand{\ind}{\mathbbm 1}
\newcommand{\Y}{\mathcal{Y}}
\newcommand{\Ucal}{\mathcal{U}}

\newcommand{\V}{\mathcal{V}}
\newcommand{\X}{\mathcal{X}}

\newcommand{\Z}{\mathcal{Z}}

\newcommand\blfootnote[1]{%
	\begingroup
	\renewcommand\thefootnote{}\footnote{#1}%
	\addtocounter{footnote}{-1}%
	\endgroup
}
\newcommand\independent{\protect\mathpalette{\protect\independenT}{\perp}}
\def\independenT#1#2{\mathrel{\rlap{$#1#2$}\mkern2mu{#1#2}}}
\definecolor{cblue}{rgb}{0.16, 0.32, 0.75}

\def\h2{\tilde h}

\def\hm1{\hat h_{-1}}

\begin{document}
\title{The Secrecy Capacity of Cost-Constrained  Wiretap Channels}
\author{Sreejith Sreekumar, Alexander Bunin,  Ziv Goldfeld, 
Haim H. Permuter, 
Shlomo Shamai (Shitz)
}

\maketitle
\begin{abstract}
In many information-theoretic channel coding problems, adding an input cost constraint to the operational setup amounts to restricting the optimization domain in the capacity formula. This paper shows that, in contrast to common belief, such a simple modification does not hold for the cost-constrained (CC) wiretap channel (WTC). The secrecy-capacity of the discrete memoryless (DM) WTC without cost constraints is described by a single auxiliary random variable. For the CC DM-WTC, however, we show that two auxiliaries are necessary to achieve capacity. Specifically, we first derive the secrecy-capacity formula, proving the direct part via superposition coding. Then, we provide an example of a CC DM-WTC whose secrecy-capacity cannot be achieved using a single auxiliary. This establishes the fundamental role of superposition coding over CC WTCs. 
\end{abstract}


\section{Introduction}

\blfootnote{The work of S. Sreekumar was supported by the TRIPODS Center for Data Science National Science Foundation Grant CCF-1740822. The work of Z. Goldfeld was supported by the National Science Foundation Grant CCF-1947801 and the 2020 IBM Academic Award. The work of H. Permuter and S. Shamai was supported by the WIN Consortium via the Israel Ministry of Economy and Science. The work of S. Shamai was also partly supported by the
European Union's Horizon 2020 Research And Innovation Programme, grant agreement no. 694630.
S. Sreekumar and Z. Goldfeld are with the School of Electrical and Computer Engineering, Cornell University, Ithaca, NY 14850, USA (email: sreejithsreekumar@cornell.edu; goldfeld@cornell.edu). A. Bunin and S. Shamai are with Department of Electrical
Engineering, Technion–Israel Institute of Technology, Haifa 3200003, Israel
(e-mail: albun@tx.technion.ac.il; sshlomo@ee.technion.ac.il). H. H. Permuter is with the Department of Electrical and Computer Engineering, Ben-Gurion University of the Negev, Beer-Sheva 8410501, Israel
(e-mail: haimp@bgu.ac.il).}

Physical-layer security (PLS), rooted in information-theoretic principles, dates back to Wyner's landmark 1975 paper \cite{Wyner_Wiretap1975}, where the wiretap channel (WTC) was introduced. This model formulates reliable and secure communication over noisy channels in the presence of an eavesdropper (see Fig. \ref{FIG:WTC}). By harnessing randomness from the noisy channel and combining it with proper physical layer coding, Wyner characterized the fundamental limit of reliable and secure communication, termed the \emph{secrecy-capacity}. For a memoryless WTC $P_{Y,Z|X}$, the secrecy-capacity is~\cite{Csiszar_Korner_BCconfidential1978} 
\begin{equation}
    C_\mathsf{\textsf{WTC}}(P_{Y,Z|X})=\max_{P_{V,X}}\, I(V;Y)-I(V;Z),\label{EQ:WTC_capacity}
\end{equation}
where the joint distribution is $P_{V,X}P_{Y,Z|X}$ (i.e., $V-X-(Y,Z)$ forms a Markov chain) and $V$ is an auxiliary random variable. PLS guarantees protection against computationally-unbounded adversaries without using shared keys. As such, it has attracted continuous attention in the information-theoretic literature, as surveyed in, e.g.,\cite{Mukherjee-2014,Wu-2018,Hamamreh-2019,Liang-book}.

In this work, we revisit the classic WTC with an input cost constraint, and show that \emph{two-layered coding is necessary} for achieving its secrecy-capacity. In many information-theoretic communication problems, adding an input cost constraint amounts to restricting the optimization domain in the capacity expression (e.g., the set of feasible $P_{V,X}$ in \eqref{EQ:WTC_capacity}). We show that this reasoning is \emph{not} valid for cost-constrained (CC) WTCs. To do so, we characterize the CC secrecy-capacity using \emph{two} auxiliary variables and prove that a single-auxiliary formula is strictly suboptimal. For the latter, an example of a CC WTC is provided for which a two-layered scheme strictly outperforms any single-layered code. This establishes the fundamental role of two-layered (superposition) coding for the CC WTC.

The necessity of two auxiliaries to achieve the secrecy-capacity of CC WTCs is perhaps surprising. This is evident from the fact that non-exact expressions for it have been used in recent works \cite[Corollary 2]{BL-2013}, \cite[Theorem 3.7]{HES-2014}. The requirement of two auxiliaries is even more remarkable when one considers the recently established analogy between WTC and Gelfand-Pinsker (GP) channels \cite{Goldfeld_WT-GP_analogy_journal2017} without cost constraints. Indeed, for the GP channel, the CC and the unconstrained capacities are given by the same expression up to adding the proper restriction to feasible input distributions \cite{GelfandPinsker1980,BarronChenWornell2003,PradhanChouRamachandran2003}. To the best of our knowledge, the WTC is the only point-to-point communication scenario for which the capacity formula itself changes due to the addition of an input cost constraint.

\subsection{Background}
The secrecy-capacity of a degraded WTC was established in \cite{Wyner_Wiretap1975}, under the so-called weak-secrecy criterion\footnote{Weak-secrecy refers to a vanishing information leakage rate $\frac{1}{n}I(M;Z^n)$ as the blocklength $n\to\infty$, where $M$ is secret message and $Z^n$ is the eavesdropper's observation.}. The secrecy-capacity of the general WTC was proved in \cite{Csiszar_Korner_BCconfidential1978}. The main result of \cite{Csiszar_Korner_BCconfidential1978} in fact accounted for a generalization of the WTC to the broadcast channel (BC) with confidential messages. The secrecy-capacity of the latter is characterized using a pair of auxiliary random variables, and the direct proof uses superposition wiretap coding. It was also shown therein that reducing the BC result to the WTC (by nullifying the common message) and requiring perfect secrecy on the private message, one of the auxiliaries can be taken to be a constant\footnote{Namely, $U$ in \cite[Corollary 2]{Csiszar_Korner_BCconfidential1978} can be taken to be a  constant $u^*$, where $u^*=\argmax_{u \in \Ucal} I(V;Y|U=u)-I(V;Z|U=u)$. As we later show, this argument cannot be applied in the CC WTC setting, as taking a constant $U$ may violate the cost constraint.}. This gave rise to the secrecy-capacity formula given in \eqref{EQ:WTC_capacity}.

Interestingly, this formula is quite robust to the security metric being used. In recent years, upgrading weak-secrecy to more stringent metrics gained much popularity. Strong-secrecy removes the normalization by the blocklength from the weak-secrecy metric, requiring that the information leakage itself vanishes \cite{Maurer_Strong_Secrecy_Chapter1994,Csiszar_Strong_Secrecy1996}. A further strengthening to the  semantic-security metric was introduced in \cite{Vardy_Semantic_WTC2012}. While weak- and strong-secrecy both assume a uniform distribution on the messages (i.e., security on average), semantic-security demands a vanishing information leakage for all probability distributions over the message set (i.e., worst-case). Despite these increasingly stringent requirement, the secrecy-capacity  with strong and semantic-security metrics remains unchanged \cite[Theorem 17.11]{Csiszar-Korner}\cite{Vardy_Semantic_WTC2012} compared to \eqref{EQ:WTC_capacity}.

In practice, transmitted signals are often bound to cost (e.g., power) constraints. Therefore, various communication scenarios originally explored without such constraints were later adapted to the CC case. This includes point-to-point channels \cite[Chapter 7]{Gallagerbook}, the GP channel \cite{BarronChenWornell2003}, and the multiple-access channel \cite[Problem 4.8]{Elgamalkim}, to name a few. Several different types of cost constraints exist in the literature such as average cost, maximal (per-codeword) cost, and   peak power constraint (for continuous alphabet input  channels).  For all these settings, the capacity under a CC is given by the same expression as in the unconstrained case, but with an added restriction on the optimization domain (see \cite{Csiszar-Korner,Han-infospectrum} and the references above).\footnote{We also mention that, apart from secrecy-capacity, other notions of the secrecy-cost trade-off exists, such as capacity per unit cost\cite{Verdu-1990,Prelov-1993,Hajek-2002,ElGamal-2006}.}
 As will be shown herein, such a simple adaptation of \eqref{EQ:WTC_capacity} to the CC case is not valid for the WTC. 
 
\subsection{Contributions}

We consider a discrete and memoryless (DM) WTC with an input cost constraint and establish a single-letter characterization of its secrecy-capacity. In contrast to \eqref{EQ:WTC_capacity}, our characterization uses two auxiliary random variables, which we show are necessary in general. As discussed above, this differentiates the WTC from other channel coding setups, where an introduction of an additional auxiliary is not needed when imposing an input cost constraint on the operational problem. We consider all three aforementioned security metrics, that is, weak-secrecy, strong-secrecy and semantic-security, and show that the secrecy-capacity is the same for them all. This is done by proving achievability under semantic-security (strongest among the three), while deriving the converse with respect to weak-secrecy.

The achievability proof uses a superposition wiretap code that carries the entire confidential message in its outer layer. The inner layer encodes only random bits purposed to confuse the eavesdropper. The cost, reliability and security analyses rely on standard random coding arguments. However, due to the presence of a cost constraint, the expected value analysis (over the codebook ensemble) does not automatically imply the existence of a deterministic codes sequence with the desired performance. We resolve this issue via a novel two-step expurgation technique that first prunes `bad' codebooks, and only then disposes of `bad' messages. A careful analysis shows that the inflicted rate loss is negligible, giving rise to a deterministic codebook that satisfies the desired cost, reliability and security requirements.

We then turn to show that two auxiliaries are necessary to achieve the CC WTC secrecy-capacity. This is done by constructing an example for which superposition coding attains a strictly higher secrecy rate than standard wiretap coding. The necessity of two auxiliaries can be understood by viewing the inner layer auxiliary as a \enquote{time-sharing} variable that leaks no information about the message to the eavesdropper. In a time-shared scheme, the cost constraint needs to be satisfied only on average (over the participating schemes). Thus, individual schemes could possibly violate the cost constraint, and indeed, it may be beneficial to consider such schemes for achieving higher secrecy rates. In particular, such a situation could occur if the mutual information term $I(V;Y)-I(V;Z)$ from the secrecy-capacity expression in \eqref{EQ:WTC_capacity} is a convex function over the CC optimization domain.

\subsection{Organization}
The remainder of this paper is organized as follows. Section \ref{Sec:Prob-setup} provides preliminary definitions and sets up the operational problem. The main results are stated and discussed in Section \ref{Sec:mainresult}, while their proofs are furnished in Section \ref{Sec:proofs}. Finally, concluding remarks are given in Section \ref{Sec:conclusion}.

\section{Preliminaries and Problem Setup} \label{Sec:Prob-setup}
\subsection{Notation}
We use the following notation. $\mathbb{N}$, $\mathbb{R}$ and $\mathbb{R}_{\geq 0}$ denotes the set of natural numbers, real numbers and non-negative real numbers, respectively. 
For  $a, b \in \mathbb{R}_{\geq 0}$, $[a:b]:=\{n \in \mathbb{N}:~a \leq n \leq b\}$. Calligraphic letters, e.g., $\mathcal{X}$, denote sets while  $|\mathcal{X}|$ stands for its cardinality. For $n \in \mathbb{N}$, $\X^n$ denotes the $n$-fold Cartesian product of $\X$, and $x^n=(x_1, \cdots,x_n)$ denotes an element of $\X^n$. Whenever the dimension $n$ is clear from the context, bold-face letters denotes vectors or sequences, e.g., $\mathbf{x}$ for $x^n$. For $ i, j \in \mathbb{N}$ such that $i \leq j$, $x_i^j:=(x_i,x_{i+1},\cdots,x_j)$; the subscript is omitted when $i=1$.

Let $(\Omega, \mathcal{F},\mathbb{P})$ be a probability space, where $\Omega$, $\mathcal{F}$ and $\mathbb{P}$ are the sample space, $\sigma$-algebra and probability measure, respectively. Random variables over $(\Omega,\mathcal{F},\mathbb{P})$ are denoted by uppercase letters, e.g., $X$, with similar conventions as above for random vectors. We use $\ind_\mathcal{A}$ for the indicator function of $\mathcal{A}\in\mathcal{F}$. The set of all probability mass functions (PMFs) on a finite set $\mathcal{X}$ (always endowed with the power set $\sigma$-algebra) is denoted by $\mathcal{P}(\mathcal{X})$.

The joint PMF of two discrete random variables $X$ and $Y$ on $(\Omega, \mathcal{F},\mathbb{P})$ is denoted by $P_{X,Y}$; the corresponding marginals are $P_X$ and $P_Y$. The conditional PMF of $X$ given $Y$ is represented by $P_{X|Y}$. Expressions such as $P_{X,Y}=P_XP_{Y|X}$ are to be understood as pointwise equality, i.e.,  $P_{X,Y}(x,y)=P_X(x)P_{Y|X}(y|x)$, for all $(x,y)\in\mathcal{X}\times\mathcal{Y}$. When the joint distribution of a triple $(X,Y,Z)$ factors as $P_{X,Y,Z}=P_{X,Y}P_{Z|Y}$, these variables form a Markov chain $X-Y-Z$. When $X$ and $Y$ are statistically independent, we write $X\independent Y$. If the entries of $X^n$ are drawn in an independent and identically distributed (i.i.d.) manner, i.e., if $P_{X^n}(x^n)=\prod_{i=1}^nP_{X}(x_i)$, $\forall~ x^n \in \X^n$, then the PMF $P_{X^n}$ is denoted by $P^{\otimes n}_X$. Similarly, if $P_{Y^n|X^n}(y^n|x^n)=\prod_{i=1}^nP_{Y|X}(y_i|x_i)$, then we write $P^{\otimes n}_{Y|X}$ for $P_{Y^n|X^n}$. The conditional product PMF given a fixed $x^n \in \X^n$ is designated by $P^{\otimes n}_{Y|X}(\cdot|x^n)$.  

For a discrete measurable space $(\mathcal{X},\mathcal{F})$, the probability measure induced by a PMF $P \in \mathcal{P}(\X)$ is denoted by $\mathbb{P}_P$; namely $\mathbb{P}_P(\mathcal{A})=\sum_{x\in\mathcal{A}}P(x)$, for all $A\in\mathcal{F}$. The corresponding expectation is designated by $\mathbb{E}_{P}$. Similarly, mutual information and entropy with an underlying PMF $P$ are denoted as $I_P$ and $H_P$, respectively. When the PMF is clear from the context, the subscript is omitted. We use $\typ{P}{\delta}{n}$ to denote the set of letter-typical sequences of length $n$ with respect to a PMF $P\in\mathcal{P}(\X)$ and a non-negative $\delta$:
\begin{align}
\mspace{-4 mu}\typ{P}{\delta}{n}\mspace{-4 mu}:= \mspace{-4 mu}\left\{\mathbf{x} \in \X^n\mspace{-4 mu}:\mspace{-4 mu} \left\vert \nu_{\mathbf{x}}(x)- \mspace{-2 mu} P(x)\right\vert \mspace{-4 mu}\leq \mspace{-4 mu}\delta P(x),~\forall x \mspace{-2 mu}\in \mspace{-2 mu}\X \right\},  
\end{align}
where $\nu_{\mathbf{x}}(x):=  \frac{1}{n} \sum_{i=1}^n\ind_{\{x_i=x\}}$ is the empirical PMF of sequence $\mathbf{x}\in\cX^n$. Finally, for a countable sample space $\cX$ and PMFs $P,Q\in\mathcal{P}(\cX)$, the Kullback-Leibler (KL) divergence between $P$ and $Q$ is
\begin{equation}
	\kl{P}{Q}:=\sum_{x\in \X}P(x)\log\left(\frac{P(x)}{Q(x)}\right),\label{EQ:relative_entropy_def_discrete}
\end{equation}
and the total variation is
\begin{align}
 \tv{P}{Q}:=\frac{1}{2}\sum_{x \in \X}\big|P(x)-Q(x)\big|.   
\end{align}
\subsection{Problem Setup}
\begin{figure}[t]
\centering
\includegraphics[trim=0cm 0cm 0cm 0cm, clip, width= 0.48\textwidth]{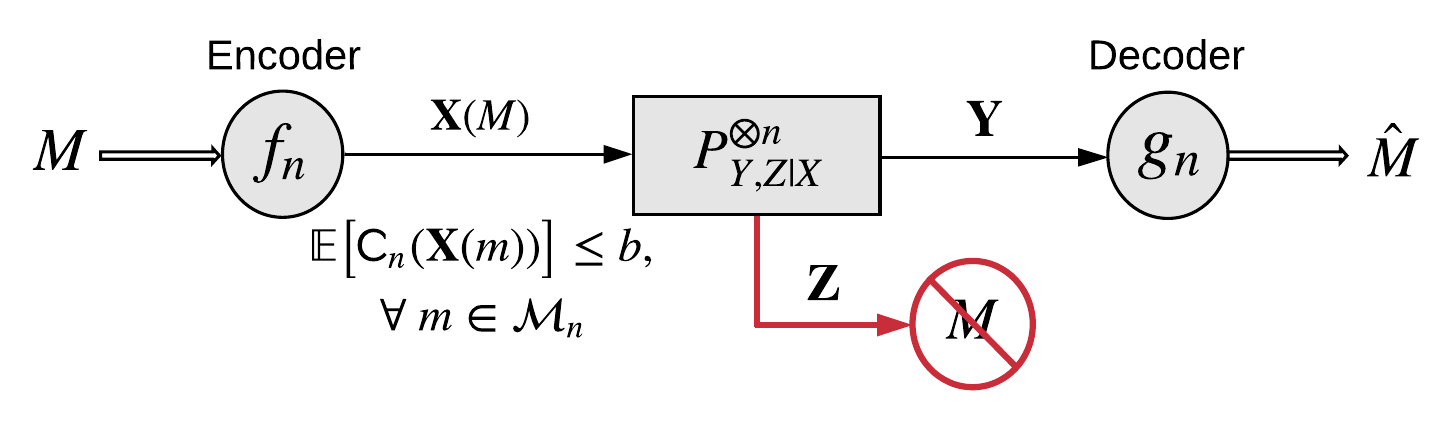}
\caption{The CC WTC  with transition kernel $P_{Y,Z|X}$. The encoder outputs $\textbf{X}(m)$ for message $M=m$, and the decoder and the eavesdropper observes $\mathbf{Y}$ and $\mathbf{Z}$, respectively. The cost constraint $\mathbb{E}\big[\mathsf{C}_n(\mathbf{X}(m))\big] \leq b$ is imposed 
for each message $m \in \mathcal{M}_n$.
} \label{FIG:WTC}
\end{figure}

Let $\X$, $\Y$ and $\Z$ be finite sets, $b \geq 0$ and $n \in \mathbb{N}$. Let $\mathsf{C}: \X \to \mathbb{R}_{\geq 0}$ be a  real-valued non-negative function. The $(\X, \Y,\Z, P_{Y,Z|X}, \mathsf{C},b)$ CC DM-WTC (henceforth referred to as CC WTC) is shown in Fig. \ref{FIG:WTC}, where $P_{Y,Z|X}:\X\to \mathcal{P}(\Y\times\Z)$ is the channel transition kernel, $\mathsf{C}$ is the cost function and $b$ is the cost constraint. The encoder chooses a message $m\in \mathcal{M}_n:=[1:2^{nR}]$, $R\geq 0$, and maps it onto a channel input sequence  $\mathbf{x} \in \X^n$. The codeword $\mathbf{x}$ is transmitted over the $n$-fold WTC $P_{Y,Z|X}^{\otimes n}$, which outputs sequences $\mathbf{y} \in \Y^n$ and $\mathbf{z} \in \Z^n$. The decoder observes $\mathbf{y}$, based on which it produces an estimate $\hat m \in \mathcal{M}_n$ of $m$. The eavesdropper observes $\mathbf{z}$, from which it tries to extract information about the transmitted message $m$. 
 
\begin{definition}[\textbf{Code}]
An $(n,R)$-code $c_n$ for a CC WTC $(\X, \Y,\Z, P_{Y,Z|X},\mathsf{C},b)$ with a message set $\mathcal{M}_n$ is a pair of functions $(f_n,g_n)$ such that
\begin{enumerate}[label = (\roman*),leftmargin=5 mm]
    \item $f_{n}:\mathcal{M}_n \to   \mathcal{P}(\X^n)$ is a stochastic encoder 
    that satisfies the per-message cost constraint given by
\begin{equation}
   \mspace{-8mu}\mathbb{E}\big[\mathsf{C}_n(\mathbf{X}(m))\big]\mspace{-5 mu}:=\mspace{-10 mu}\sum_{\mathbf{x}\in\X^n}\mspace{-10 mu}f_n(\mathbf{x}|m)\mathsf{C}_n(\mathbf{x}) \leq b,~\forall~ m \in \mspace{-3 mu}\mathcal{M}_n, \label{permsgcost}\\
\end{equation}
where, $\mathbf{X}(m)\sim f_n(\cdot|m)$, and $\mathsf{C}_n(\mathbf{x}):= \frac{1}{n}\sum_{i=1}^n \mathsf{C}(x_i)$, $\forall ~\mathbf{x}\in \X^n$, is the $n$-fold extension of $\mathsf{C}$;
    \item $g_n: \Y^n \to \mathcal{M}_n$ is the decoding function.
\end{enumerate}
\end{definition}
\begin{remark}[\textbf{Minimal cost}]
The per-message constraint in \eqref{permsgcost} can be satisfied only if  $b \geq \mathsf{c}_{\min}:=\min\{\mathsf{C}(x): x \in \X\}$. Henceforth, we will assume this condition holds without further mention.
\end{remark}
A message PMF $P_{M}\in\cP(\cM_n)$ and a code $c_n=(f_n,g_n)$ induces a PMF on $\cM_n\times\cX^n\times\cY^n\times\cZ^n\times\cM_n$ given by
\begin{align}
    P^{(c_n)}(m,\mathbf{x}, \mathbf{y},\mathbf{z}, \hat m)&=P_M(m)~f_{n}(\mathbf{x}|m)~P^{\otimes n}_{Y,Z|X}(\mathbf{y},\mathbf{z}|\mathbf{x}) \notag \\
    &\qquad \ind_{\{\hat m=g_n(\mathbf{y})\}}.
\end{align}
The performance of $c_n$ is evaluated in terms of the maximal decoding error probability and a chosen security metric.
\begin{definition}[\textbf{Error Probability}]
The maximal error probability of an $(n,R)$-code $c_n$ is  
\begin{subequations}
\begin{equation}
   e(c_n):=\max_{m \in \mathcal{M}_n} e_m(c_n), \label{maxerrpconst}
\end{equation}
where for any $m \in \mathcal{M}_n$, 
\begin{align}
    e_m(c_n)&:=\mathbb{P}_{P^{(c_n)}}\big(\hat M \neq m\big|M=m\big) \notag \\ &=\sum_{\mathbf{x} \in \X^n} f_{n}(\mathbf{x}|m)\sum_{\substack{\mathbf{y} \in \Y^n:\\g_n(\mathbf{y}) \neq m}}P_{Y|X}^{\otimes n}(\mathbf{y}|\mathbf{x}).\label{err-prob1}
\end{align}
\end{subequations}
\end{definition}
\begin{definition}[\textbf{Security metrics}]
For a given  $(n,R)$-code $c_n$ and message PMF $P_M$, the information leakage to the eavesdropper is $\ell(P_M,c_n):=I_{P^{(c_n)}}(M;\mathbf{Z})$. The weak-secrecy, strong-secrecy and semantic-security metrics with respect to (w.r.t.) the code $c_n$ are, respectively, defined as
\begin{align}
\ell_\mathsf{weak}(c_n)&:= \frac{1}{n}\ell(\bar {P}_M,c_n)\\
\ell_\mathsf{str}(c_n)&:= \ell(\bar {P}_M,c_n) \\
\ell_\mathsf{sem}(c_n)&:=\max_{P_M \in \mathcal{P}(\mathcal{M}_n) } \ell(P_M,c_n),
\end{align}
where $\bar P_M$ denotes the uniform distribution on $\mathcal{M}_n$.
\end{definition}
\begin{definition}[\textbf{Achievability}]
A rate $R \geq 0$ is said to be achievable with semantic-security for a  CC WTC $(\X, \Y,\Z, P_{Y,Z|X},\mathsf{C}, b)$ if for every $\epsilon>0$ and sufficiently large $n$, there exists an $(n,R)$-code $c_n$ such that \eqref{permsgcost} is satisfied, and
\begin{align}
    \max \big\{e(c_n),\ell_\mathsf{sem}(c_n)\big\} \leq \epsilon. \label{relsecconst}
\end{align}
Achievability w.r.t. the weak- or  strong-secrecy metrics is defined by replacing $\ell_\mathsf{sem}$ with $\ell_\mathsf{weak}$ or $\ell_\mathsf{str}$, respectively.
\end{definition}
\begin{definition}[\textbf{Secrecy Capacity}]
The semantic-security capacity of a CC WTC $(\X, \Y,\Z, P_{Y,Z|X},\mathsf{C}, b)$, denoted by $C_\mathsf{sem}(P_{Y,Z|X},\mathsf{C},b)$, is the supremum of the set of all rates achievable with semantic-security. The strong-secrecy-capacity $C_\mathsf{str}(P_{Y,Z|X},\mathsf{C},b)$ and weak-secrecy-capacity $C_\mathsf{weak}(P_{Y,Z|X},\mathsf{C},b)$ are defined similarly w.r.t. the corresponding notion of achievability.
\end{definition}


\section{Main Results} \label{Sec:mainresult}
We give a single-letter characterization of the weak-secrecy, strong-secrecy and semantic-security capacities of the CC WTC, all of which are shown to be equal. The characterization involves two auxiliary random variables. Both auxiliaries are necessary to achieve capacity in general. We first state the capacity result, and then provide an example for which any single-auxiliary scheme is suboptimal.

\subsection{Secrecy-capacity results}

Let $\cU$ and $\cV$ be finite sets, and for any $P_{U,V,X}\in\cP(\cU\times\cV\times\cX)$, set
\begin{equation}
    \tilde C(P_{U,V,X},P_{Y,Z|X}):=I_P(V;Y|U)-I_P(V;Z|U), \label{sscapexpint}
\end{equation}
where the mutual information terms are taken w.r.t. $P=P_{U,V,X}P_{Y,Z|X}$. Also, let
\begin{flalign} \label{setfeasdist}
&\mathcal{H}(\mathsf{C},b)\notag \\
&:= \left\{\mspace{-3 mu}P_{U,V,X}\mspace{-3 mu}\in\mspace{-3 mu}\cP(\cU\times\mspace{-3 mu}\cV\mspace{-3 mu}\times\mspace{-3 mu}\cX)\mspace{-3 mu}:\begin{aligned} & \mspace{-3 mu}P_{U,V,X}\mspace{-3 mu}=\mspace{-3 mu}P_{U,V}\mspace{-2 mu}P_{X|V},\\ &\mspace{-3 mu} \mathbb{E}_P\big[\mathsf{C}(X)\big] \leq b\end{aligned}\right\},
\end{flalign}
and define
\begin{equation} \label{sscapexp}
    \bar  C(P_{Y,Z|X},\mathsf{C},b):=
 \sup_{P_{U,V,X}\in\cH(\mathsf{C},b)}~\tilde C(P_{Y,Z|X},P_{U,V,X}).
\end{equation}

Our main result is given next. For simplicity of presentation, we will suppress $P_{Y,Z|X}$ and $\mathsf{C}$ from the notation $C_\mathsf{sem}(P_{Y,Z|X},\mathsf{C},b)$, $C_\mathsf{str}(P_{Y,Z|X},\mathsf{C},b)$, $C_\mathsf{weak}(P_{Y,Z|X},\mathsf{C},b)$, $  \tilde C(P_{Y,Z|X},P_{U,V,X})$, $ \mathcal{H}(\mathsf{C},b)$ and $\bar C(P_{Y,Z|X},\mathsf{C},b)$, henceforth denoting them by $C_\mathsf{sem}(b)$, $C_\mathsf{str}(b)$, $C_\mathsf{weak}(b)$, $  \tilde C(P_{U,V,X})$, $ \mathcal{H}(b)$ and $\bar C(b)$, respectively.
\begin{theorem}[\textbf{Secrecy-capacity}] \label{thm:main}
The secrecy-capacity of a CC WTC $(\X, \Y,\Z, P_{Y,Z|X},\mathsf{C}, b)$ under weak-secrecy, strong-secrecy and semantic-security is the same, and is given by
\begin{align}
    C_\mathsf{sem}(b)=C_\mathsf{str}(b)=C_\mathsf{weak}(b)=\bar C(b).\label{EQ:main}
\end{align}
\end{theorem}
The proof of Theorem \ref{thm:main} is given in Section \ref{Sec:mainproof}. 
The achievability of \eqref{EQ:main} relies on a superposition wiretap coding, while the converse adapts the classic WTC converse to accommodate the cost constraint. We note that the CC WTC's secrecy-capacity expression involves two auxiliary random variables $U$ and $V$. In contrast, the secrecy-capacity formula of a WTC (without a cost constraint), given in \eqref{EQ:WTC_capacity}, uses only a single auxiliary. In Section \ref{SUBSEC:1v2_aux} we show that a reduction of $\bar C(b)$ to a single auxiliary is impossible, in general.
\begin{remark}[\textbf{Per-codeword cost constraint}] \label{remarkpcconst}
In addition to the per-message cost constraint in \eqref{permsgcost}, Theorem \ref{thm:main} can be extended to the more stringent scenario of a per-codeword cost constraint, i.e., 
\begin{flalign}
 &\mathsf{C}_n(\mathbf{x})\mspace{-3 mu} \leq b,\forall~\mspace{-3 mu}\mathbf{x}\mspace{-3 mu} \in \mspace{-3 mu}\X^n \mbox{ s.t. }\exists~\mspace{-3 mu} m \mspace{-3 mu} \in \mspace{-3 mu} \cM_n \mbox{ with } f_n(\mathbf{x}|m)\mspace{-2 mu}>\mspace{-2 mu}0. \label{percodewrdconst} 
\end{flalign}
In other words, re-defining achievability by replacing \eqref{permsgcost} with \eqref{percodewrdconst}, the same $\bar C(b)$ is obtained as the per-codeword cost secrecy-capacity under weak-secrecy, strong-secrecy and semantic security. The argument relies on a natural modification of the achievability proof of Theorem \ref{thm:main}, by employing constant-composition superposition wiretap codes instead of the currently used i.i.d. ensemble (the converse follows from the per-message cost case). A central  ingredient for the modified security analysis is a version of Lemma \ref{Lem:SSCL} for the constant-composition ensemble, which can be obtained by extending the results in \cite{Yagli-2019} to superposition codes.
\end{remark}
The following lemma provides additional properties of $\bar C(b)$. These properties are used in the proof of Theorem~\ref{thm:main}. 
\begin{lemma}[\textbf{Structural properties}]\label{lem:concave}
In the definition of $\bar C(b)$ in \eqref{sscapexp}, it suffices to consider auxiliary alphabets $\cU$ and $\cV$ with $|\cU| \leq |\cX|$ and $|\cV| \leq |\cX|^2$. Moreover, $\bar C(b)$ is a non-decreasing and  concave (for $b \geq \mathsf{c}_{\min}$)  function of $b$
, and the supremum in \eqref{sscapexp} is achieved, i.e., 
\begin{align}
      \bar C(b)&=\max_{P_{U,V,X}\in\cH(b)}\tilde C(P_{U,V,X})\notag\\
      &:=\max_{P_{U,V,X}\in\cH(b)}I_P(V;Y|U)-I_P(V;Z|U).\label{SScapmaxexp}
\end{align}
 \end{lemma}
The proof of Lemma \ref{lem:concave} is provided in Appendix \ref{App1} for completeness.
\subsection{Two Auxiliaries are Necessary}\label{SUBSEC:1v2_aux}
Comparing \eqref{EQ:WTC_capacity} and \eqref{SScapmaxexp}, one might ask whether a reduction to a single auxiliary random variable in Theorem \ref{thm:main} is possible. We show that the answer is negative in general. To this end, we provide an example of a CC WTC $P_{Y,Z|X}$, for which
\begin{align}
   & \max_{\substack{P_{U,V,X}\in \mathcal{H}(b)} } I_P(V;Y|U)-I_P(V;Z|U) \notag \\
   &\qquad \qquad \qquad > \max_{\substack{P_{V,X}:\\\mathbb{E}_P[\mathsf{C}(X)] \leq b}}    I_P(V;Y)-I_P(V;Z), \label{twoauxreqd}
\end{align}
where the mutual information terms on the right hand side (RHS) are w.r.t. $P_{V,X}P_{Y,Z|X}$. To explain briefly, the example incorporates a WTC setup in which the transmitter (encoder) is connected to the receiver (decoder) via a noiseless private data link. The transmitter can choose the content as well as timing of the transmission,  however, it is constrained to use the link at most half of the time. 
The receiver observes the data (error-free)  when transmission occurs, and  random noise otherwise. On the other hand, the eavesdropper has no access to the data link, but perfectly knows the timing of the transmission. We next describe the details of the WTC setup.

\begin{figure}[t]
\centering
\includegraphics[trim=0.6cm 0cm 0cm 0cm, clip, width= 0.5\textwidth]{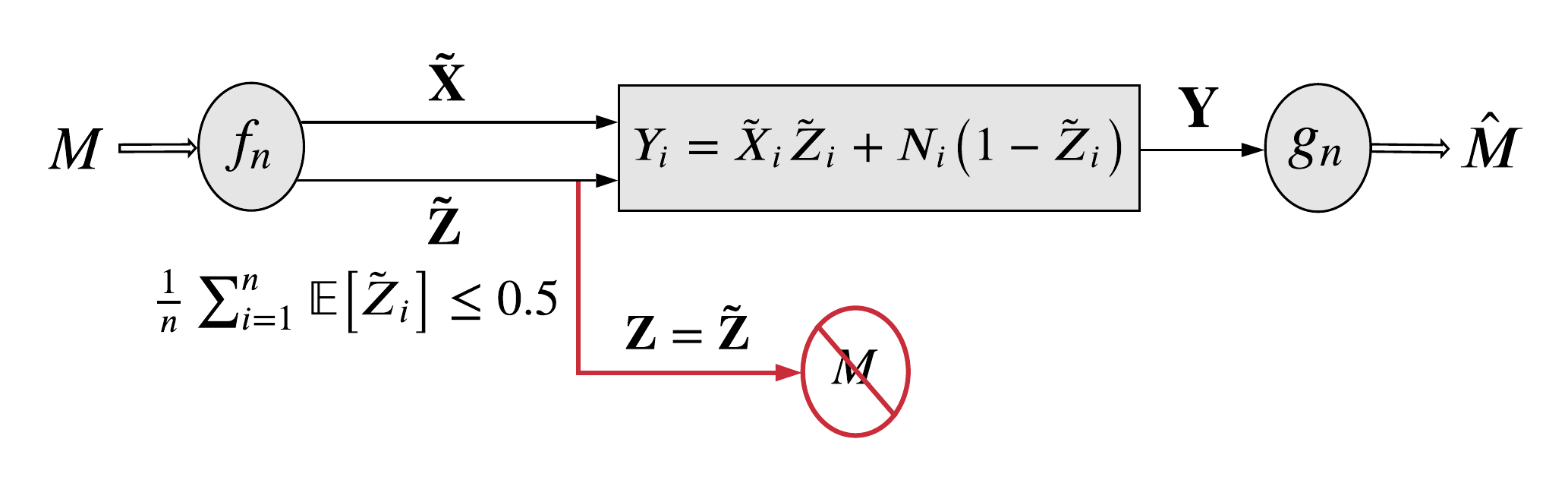}
\caption{
The CC WTC  with transition kernel $P_{Y,Z|X}$ used in Proposition \ref{Prop:example}. The encoder outputs $\mathbf{X}=(\tilde{\mathbf{X}}, \mathbf{\tilde{Z}})$ such that the cost constraint $\frac{1}{n} \sum_{i=1}^n\mathbb{E}(\tilde Z_i) \leq 0.5$ is satisfied. The eavesdropper's observation $\mathbf{Z}=\mathbf{\tilde{Z}}$ is  controlled by the encoder, and the realization of $\tilde Z_i$ decides whether the decoder observes $\tilde{X}_i$ (when $\tilde Z_i=1$) or noise $N_i$ (when $\tilde Z_i=0$). 
} \label{FIG:WTC-Example}
\end{figure}

Consider the $(\X, \Y,\Z, P_{Y,Z|X},\mathsf{C}, b)$ CC WTC shown in Fig. \ref{FIG:WTC-Example} that is defined as follows:
\begin{itemize}
\item Let $\tilde{\X}=\Y=\Z=\{0,1\}$, and $\X=\tilde{\X} \times \Z$.
\item The channel input is $X=(\tilde X,\tilde Z)\sim P_{\tilde X,\tilde Z}$, where $\tilde X$ and  $\tilde Z$ take values in $\tilde{\mathcal{X}}$ and $\cZ$, respectively, and both are controlled by the encoder.
\item Consider the cost function $\mathsf{C}(x)=\mathsf{C}(\tilde x,\tilde z)=\tilde z$, for all $x=(\tilde x,\tilde z) \in \tilde{\X} \times \Z$, and set the cost constraint to $b=0.5$. Thus, the input must satisfy $$\mathbb{E}\big[\mathsf{C}(X)\big]=\mathbb{E}\big[\tilde Z\big] \leq \frac{1}{2}.$$
\item Let $N\sim \textsf{Ber} (0.5)$ be independent of $X=(\tilde X,\tilde Z)$ and set $$Y=\tilde X \tilde Z+N(1-\tilde Z).$$
Let $P_{Y|X}=P_{Y|\tilde X,\tilde Z}$ denote the transition kernel from $X$ to $Y$ induced by the above relation.
\item The WTC $P_{Y,Z|X}$ is given by $P_{Y,Z|\tilde X,\tilde Z}=P_{Y|\tilde X,\tilde Z}\ind_{\{Z=\tilde Z\}}$.
\end{itemize}

We have the following proposition whose proof is given in Section \ref{Sec:Proof:Example}.
\begin{prop}[\textbf{Necessity of two auxiliaries}] \label{Prop:example}
For the $(\X, \Y,\Z, P_{Y,Z|X},\mathsf{C},0.5)$ CC WTC described above, 
\begin{align}
 \bar C(0.5)&:=\max_{\substack{P_{U,V,X}\in \mathcal{H}(0.5)} } I_P(V;Y|U)-I_P(V;Z|U) \notag \\
 &\geq 0.5> \max_{\substack{P_{V,X}:\\\mathbb{E}_P[\mathsf{C}(X)] \leq 0.5}}    I_P(V;Y)-I_P(V;Z). \label{singauxvstwoaux}
\end{align}
\end{prop}
For proving Proposition \ref{Prop:example}, we choose a $P_{U,V,X} \in \mathcal{H}(0.5)$ such that $I_P(V;Y|U)-I_P(V;Z|U)  = 0.5$, thus establishing  $\bar C(0.5) \geq 0.5$. This is done by selecting $U=\tilde Z \sim \textsf{Ber}(0.5)$, $\tilde X\sim \textsf{Ber}(0.5) \independent \tilde Z$,  and $V=X:=(\tilde X, \tilde Z)$. Intuitively, such a choice of auxiliaries correspond to a communication scheme of transmitting data half  of the time (say, every alternate channel use). Subsequently, we show that the RHS of \eqref{singauxvstwoaux} is strictly below $0.5$. This is shown by starting with the assumption that there exists a $P_{V,X}$ such that $\mathbb{E}_P[\mathsf{C}(X)] \leq 0.5$ and $ I_P(V;Y)-I_P(V;Z) \geq 0.5$, and then arguing that it leads to a contradiction. 

Achieving the CC WTC secrecy-capacity in the above example requires two auxiliary random variables. While a reduction in the number of auxiliaries is impossible in general, we next show that no auxiliaries are needed when the WTC is less noisy. Namely, the condition is that $Y$ is \textit{less noisy} than $Z$, i.e., $I_P(U;Y) \geq I_P(U;Z)$, for all $P_{U,X,Y,Z}=P_{U,X}P_{Y,Z|X}$ \cite{Korner-Marton-1977}. This is similar to the state of affairs for WTC without a cost constraint\cite{Csiszar_Korner_BCconfidential1978}.
\begin{cor}[\textbf{Less noisy WTCs}] \label{Cor:lessnoisy}If $Y$ is less noisy than $Z$, then 
\begin{align} \label{MCseccap}
      \bar C(b) = 
   \max_{P_X:~ \mathbb{E}_P[\mathsf{C}(X)]\leq b}~ I_P(X;Y)-I_P(X;Z).
\end{align}
If $Z$ is less noisy than $Y$, then $\bar C(b)=0$.
\end{cor}
The proof of Corollary \ref{Cor:lessnoisy} is provided in Section \ref{Cor:proof}.
\section{Proofs}
\label{Sec:proofs}
\subsection{Proof of  Theorem \ref{thm:main}} \label{Sec:mainproof}
  We will show that $C_\mathsf{weak}(b) \leq  \bar C(b)$ and $C_\mathsf{sem}(b) \geq \bar C(b)$ for any $b \geq 0$. This combined with the fact that $C_\mathsf{sem}(b) \leq C_\mathsf{str}(b) \leq C_\mathsf{weak}(b)$ will imply the desired result.
 \subsubsection{\underline{Converse}} 
 Recall that $\bar P_M$ denotes the uniform distribution on $\mathcal{M}_n$. It suffices to show that for any $\delta>0$, a rate $R$ achievable under the weak-secrecy metric satisfies $R \leq \bar C(b)+\delta$, for large enough $n$. Fix $\epsilon>0$ and let $c_n$ be an $(n,R)$-code with $\max \big\{e(c_n),\ell_\mathsf{weak}(c_n)\big\} \leq \epsilon$. Any such code must also satisfy the weaker constraint $\max \left\lbrace\mathbb{E}_{\bar P_M}\big[e_M(c_n)\big],\frac{1}{n}\ell(\bar P_M, c_n)\right\rbrace \leq \epsilon$. Accordingly, we may assume without loss of generality 
 that $P_M=\bar P_M$. From Fano's inequality \cite{CoverThomas}, it follows that
 \begin{align}
     H(M|\mathbf{Y}) \leq 1+\epsilon nR . \label{Eq:fanoineqconseq}
 \end{align}
Now, we can write
 \begin{flalign}
    &nR \notag \\
    &=H(M) \notag \\
     &\stackrel{(a)}{\leq} I(M;\mathbf{Y})+1+\epsilon nR \notag \\
     & \stackrel{(b)}{\leq} I(M;\mathbf{Y})-I(M;\mathbf{Z})+1+\epsilon nR+n\epsilon \notag \\
     &=\sum_{i=1}^n I(M;Y_i|Y^{i-1})-I(M;Z_i|Z_{i+1}^n)+1+\epsilon n(R+1) \notag \\
    &\stackrel{(c)}{=}\sum_{i=1}^n I(M;Y_i|Y^{i-1})-I(M;Z_i|Z_{i+1}^n)+\epsilon n(R+1)\notag \\
    &\qquad \quad +I(Z_{i+1}^n;Y_i|M,Y^{i-1})-I(Y^{i-1};Z_i|M,Z_{i+1}^n)+1\notag \\
     &=\sum_{i=1}^n I(M,Z_{i+1}^n;Y_i|Y^{i-1})-I(M,Y^{i-1};Z_i|Z_{i+1}^n)\notag \\
       &\qquad \quad+1+\epsilon n(R+1) \notag   \\
     &\stackrel{(d)}{=}\sum_{i=1}^n I(M;Y_i|Y^{i-1},Z_{i+1}^n)-I(M;Z_i|Y^{i-1},Z_{i+1}^n)\notag \\
       &\qquad \quad+1+\epsilon n(R+1) \notag   \\
 &=\sum_{i=1}^n I(M,Y^{i-1},Z_{i+1}^n;Y_i|Y^{i-1},Z_{i+1}^n)+1+\epsilon n(R+1)\notag \\ &\qquad \quad-I(M,Y^{i-1},Z_{i+1}^n;Z_i|Y^{i-1},Z_{i+1}^n)\notag  \\
 &\stackrel{(e)}{=}\sum_{i=1}^n I(V_i;Y_i|U_i)-I(V_i;Z_i|U_i)+1+\epsilon n(R+1)\notag \\
     & \stackrel{(f)}{\leq}  \sum_{i=1}^n  \bar C\Big(\mathbb E\big[\mathsf{C}(X_i)\big] \Big)+1+\epsilon n(R+1) \notag \\
     & \stackrel{(g)}{\leq}  n \bar C\left(\sum_{i=1}^n \frac 1n\mathbb E\big[\mathsf{C}(X_i)\big]\right) +1+\epsilon n(R+1) \notag \\
     &\stackrel{(h)}{=} n \left(\bar C(b) + \epsilon (R+1)+\frac{1}{n} \right), \notag &&
  \end{flalign} 
where
\begin{enumerate}[label = (\alph*),leftmargin=5.5mm]
    \item follows from \eqref{Eq:fanoineqconseq};
    \item is because $l(\bar P_M,c_n) \leq n\epsilon$;
    \item and (d) use the  Csisz\'{a}r-sum identity \cite{Elgamalkim}; \addtocounter{enumi}{1}
    \item  is due to the auxiliary random variable identification $U_i=(Y^{i-1},Z_{i+1}^n)$ and $V_i=(M,Y^{i-1},Z_{i+1}^n)$;
    \item follows from the definition of $\bar C(\cdot)$ in \eqref{sscapexp} since $U_i-V_i-X_i-(Y_i,Z_i)$ form a Markov chain with the above auxiliary variable identification;
    \item is due to the concavity of $\bar C(\cdot)$  proved in Lemma \ref{lem:concave};
\item  is because $\bar C(\cdot)$ is non-decreasing and  $X^n$ satisfies $\mathbb E \Big[\frac{1}{n}\sum_{i=1}^n \mathsf{C}(X_i)\Big] \leq b$ due to \eqref{permsgcost}.  
\end{enumerate}
Thus, 
\begin{align}
    R \leq \frac{\bar C(b) + \epsilon+\frac{1}{n}}{1-\epsilon}. \notag
\end{align}
The claim follows by taking $n$ sufficiently large and $\epsilon>0$ small enough.

\vspace{2mm}
\subsubsection{\underline{Achievability}} By the continuity\footnote{This follows from the concavity of $\bar C(b)$; 
see Lemma \ref{lem:concave}.} of $\bar C(b)$, 
it suffices to show that for any $b>\mathsf{c}_{\min}$ and  $\epsilon>0$, 
there exists an $(n,R)$ code $c_n$ that satisfies \eqref{relsecconst}, provided that $R < \bar C(b)$ and $n$ is sufficiently large. To this end, we construct an ensemble of superposition wiretap codes and show that the expected (over the ensemble) cost, error probability and  semantic-security metric satisfy average versions of the constraints. Then, through a sequence of codebook and message expurgation steps, we show the existence of a code $c_n$ that satisfies \eqref{permsgcost} and \eqref{relsecconst}, as required.

Fix $\epsilon>0$ and a joint  PMF $P_{U,V,X,Y,Z}:=P_{U,V}P_{X|V}P_{Y,Z|X} \in \mathcal{P}(\Ucal \times \V \times \X \times \Y \times \Z)$ such that $\mathbb{E}_P\big[\mathsf{C}(X)\big]< b$. \\
\textbf{Codebook $\bm{\mathcal{B}_{n}}$:} We use a superposition codebook such that the inner layer adds redundancy to confuse the eavesdropper, while the outer layer carries the information about the message.

Define the index sets $\mathcal{I}_n:=[1:2^{nR_1}]$ and $\mathcal{J}_n:=[1:2^{nR_2}]$. Let $\mathbb{B}_{U}^{(n)}:=\{\mathbf{U}(i),~ i \in \mathcal{I}_n\}$ be a random inner layer codebook such that each codeword $\mathbf{U}(i),~ i \in \mathcal{I}_n$, is a sequence of length $n$ generated independently according to $P^{\otimes n}_U$. Denote a realization of $\mathbb{B}_{U}^{(n)}$ by $\mathcal{B}_{U}^{(n)}:=\{\mathbf{u}(i),~ i \in \mathcal{I}_n\}$. 

For a fixed $\mathcal{B}_{U}^{(n)}$ and each $i \in \mathcal{I}_n$, let $\mathbb{B}_V^{(n)}(i):=\{\mathbf{V}(i,j,m), (j,m) \in \mathcal{J}_n \times \mathcal{M}_n\}$ denote a collection of $n$-length random sequences, each generated  independently according to $P^{\otimes n}_{V|U}\big(\cdot\big|\mathbf{u}(i)\big)$. Denote a realization of $\mathbb{B}_V^{(n)}(i)$ by $\mathcal{B}_V^{(n)}(i):=\{\mathbf{v}(i,j,m), (j,m) \in \mathcal{J}_n \times \mathcal{M}_n\}$. Also, set $\mathbb{B}_V^{(n)}:=\{\mathbb{B}_V^{(n)}(i),~i \in \mathcal{I}_n\}$, and denote its possible outcome by $\mathcal{B}_V^{(n)}$. With the above, the random superposition codebook is given by $\mathbb{B}_{n}:=\{\mathbb{B}_U^{(n)}, \mathbb{B}_V^{(n)}\}$, and its realization is denoted by~$\mathcal{B}_{n}$.

Denoting the set of all possible values of $\mathbb{B}_{n}$ by $\mathfrak{B}_{n}$, the codebook construction described above induces a PMF $\mu\in\cP(\mathfrak{B}_{n})$, given by
\begin{flalign}
   & \mu(\mathcal{B}_{n}):=\mu\left(\mathcal{B}_U^{(n)}, \mathcal{B}_V^{(n)}\right)\notag \\
    &:=\prod_{i \in \mathcal{I}_n} \left[P^{\otimes n}_{U}\big(\mathbf{u}(i)\big) \prod_{(j,m) \in  \mathcal{J}_n \times \mathcal{M}_n} \mspace{-10 mu} P^{\otimes n}_{V|U}\big(\mathbf{v}(i,j,m)\big|\mathbf{u}(i)\big)\right]. \notag &&
\end{flalign}

\textbf{Encoder $\bm{f_n}$:} Given a codebook $\mathcal{B}_{n}$ and message $M=m$, the encoder chooses an index pair $(i,j)$ uniformly at random from the set $\mathcal{I}_n \times \mathcal{J}_n$, and transmits  $\mathbf{X}\sim P^{\otimes n}_{\mathbf{X}|\mathbf{V}}\big(\cdot\big|\mathbf{v}(i,j,m)\big)$. 
The induced encoding function $f_n:\cM_n\to\cP(\cX^n)$ is given 
\begin{align}
    f_n(\mathbf{x}|m)& =\frac{1}{|\cI_n||\cJ_n|}\sum_{(i,j)\in\cI_n\times\cJ_n} P^{\otimes n}_{\mathbf{X}|\mathbf{V}}\big(\mathbf{x}\big|\mathbf{v}(i,j,m)\big), \notag \\
    &\qquad \forall~ (m,\mathbf{x})\in\cM_n\times \cX^n. \label{eq:encodingfn}
\end{align}

\textbf{Decoder $\bm{g_n}$:} Upon observing $\mathbf{y} \in \Y^n$, the decoder looks for a unique tuple 
$\big(\hat i,\hat j,\hat m\big) \in \mathcal{I}_n \times \mathcal{J}_n \times \mathcal{M}_n$ such that $\Big(\mathbf{u}\big(\hat i), \mathbf{v}(\hat i,\hat j,\hat m\big),\mathbf{y}\Big) \in \typ{P_{U,V,Y}}{\delta}{n}$, for some  $\delta>0$. 
If such a unique tuple exists, the decoder sets $g_n(\mathbf{y})=\hat m$; else, $g_n(\mathbf{y})=1$.

\textbf{Induced distribution:} Denote the pair $(f_{n},g_n)$ w.r.t. the codebook $\mathcal{B}_{n}$ by $c_n(\mathcal{B}_{n})$. For a given codebook $\mathcal{B}_{n}$, $c_n(\mathcal{B}_{n})$ induces a joint PMF $P^{(\mathcal{B}_{n})}_{M,I,J,\mathbf{U},\mathbf{V},\mathbf{X},\mathbf{Y},\mathbf{Z},\hat M} \in \mathcal{P}(\mathcal{M}_n \times \mathcal{I}_n \times \mathcal{J}_n \times \Ucal^n \times \V^n \times \X^n \times \Y^n \times \Z^n \times  \mathcal{M}_n)$, given~by
\begin{flalign}
  &  P^{(\mathcal{B}_{n})}_{M,I,J,\mathbf{U},\mathbf{V},\mathbf{X},\mathbf{Y},\mathbf{Z},\hat M}(m,i,j,\mathbf{u},\mathbf{v},\mathbf{x},\mathbf{y},\mathbf{z}, \hat m)\notag\\
    &= P_M(m) \frac{1}{|\cI_n||\cJ_n|} \ind_{\{\mathbf{u}=\mathbf{u}(i),\mathbf{v}=\mathbf{v}(i,j,m)\}}P^{\otimes n}_{X|V}(\mathbf{x}|\mathbf{v})\notag \\
    &\qquad P^{\otimes n}_{YZ|X}(\mathbf{y},\mathbf{z}|\mathbf{x}) \ind_{\{\hat m=g_n(\mathbf{y})\}}.
\end{flalign}
Henceforth, we use $P^{(\mathcal{B}_{n})}$ and $P^{(\mathbb{B}_{n})}$ as shorthands for 
$P^{(\mathcal{B}_{n})}_{M,I,J,\mathbf{U},\mathbf{V},\mathbf{X},\mathbf{Y},\mathbf{Z},\hat M}$ and $P^{(\mathbb{B}_{n})}_{M,I,J,\mathbf{U},\mathbf{V},\mathbf{X},\mathbf{Y},\mathbf{Z},\hat M}$, respectively.  We will also  denote the probability measure induced by $P^{(\mathcal{B}_{n})}$ and $P^{(\mathbb{B}_{n})}$  by $\mathbb{P}_{P^{(\mathcal{B}_{n})}}$ and $\mathbb{P}_{P^{(\mathbb{B}_{n})}}$, respectively. Note that  $P^{(\mathbb{B}_{n})}$ is a random PMF and $\mathbb{P}_{P^{(\mathbb{B}_{n})}}$ is a random probability measure.

\textbf{Cost Analysis:} We analyze the expectation (w.r.t. the random codebook) of the cost averaged over messages. For any $m \in \mathcal{M}_n$, note that $\mathbb{E}_{\mu}\Big[ P_{\mathbf{X}|M}^{(\mathbb{B}_{n})}(\mathbf{x}|m)\Big]=P^{\otimes n}_X(\mathbf{x})$, $\forall~ \mathbf{x} \in \X^n$, which readily implies that
\begin{align}
    \mathbb{E}_{\mu} \Big[ \mathbb{E}_{P_{\mathbf{X}|M}^{(\mathbb{B}_{n})}(\cdot|m)}\big[\mathsf{C}_n(\mathbf{X})\big]\Big] &= \mathbb{E}_{P^{\otimes n}_X}\big[\mathsf{C}_n(\mathbf{X})\big]\notag \\
    &=\mathbb{E}_{P_X}\big[\mathsf{C}(X)\big] < b. \notag
\end{align}
 It follows that for some $\gamma'>0$ and all $n \in \mathbb{N}$,
\begin{align}
    \mathbb{E}_{\mu} \Big[ \mathbb{E}_{P_{\mathbf{X}}^{(\mathbb{B}_{n})}}\big[\mathsf{C}_n(\mathbf{X})\big]\Big] \leq b-\gamma'. \label{costavg}
\end{align}

\textbf{Average error probability analysis:} 
We analyze the expected error probability averaged over messages. For any $\cB_n\in\mathfrak{B}_n$ and $(i,j,m)\in\cI_n\times\cJ_n\times\cM_n$, let $\mathbf{Y}\sim P^{(\cB_n)}_{\mathbf{Y}|I,J,M}(\cdot|i,j,m)$, and define the following error events:
\begin{align*} \mathcal{E}_1(i,j,m)&:=\Big\{\big(\mathbf{u}(i),\mathbf{v}(i,j,m),\mathbf{Y}\big) \notin \typ{P_{U,V,Y}}{\delta}{n}\Big\}\\
\mathcal{E}_2(i,j,m)&:=\Big\{\exists(j',m') \in \mathcal{J}_n \times \mathcal{M}_n,~\big(j',m'\big) \neq (j,m):\notag \\
& \qquad \big(\mathbf{u}(i),\mathbf{v}(i,j',m'),\mathbf{Y}\big) \in \typ{P_{U,V,Y}}{\delta}{n}\Big\}\\
\mathcal{E}_3(i,j,m)&:=\Big\{\exists\big(i',j',m'\big)\mspace{-1mu} \in \mathcal{I}_n\mspace{-1mu} \times\mspace{-1mu} \mathcal{J}_n\mspace{-1mu} \times\mspace{-1mu} \mathcal{M}_n,~i'\mspace{-2mu} \neq\mspace{-2mu} i:\notag \\
& \qquad \big(\mathbf{u}(i'),\mathbf{v}\big(i',j',m'\big),\mathbf{Y}\big)\in \typ{P_{U,V,Y}}{\delta}{n}\Big\}.
\end{align*}
Due to the symmetry of the random codebook $\mathbb{B}_{n}$, encoder $f_{n}$ and decoder $g_n$, the expected error probability over $\mathbb{B}_{n}$, i.e.,  $\mathbb{E}_{\mu}\Big[ \mathbb{P}_{P^{(\mathbb{B}_{n})}}(\hat M \neq M)\Big]$, is the same for any realization of $(I,J,M)$. Thus, we may assume without loss of generality 
that $(I,J,M)=(1,1,1)$. We have
\begin{flalign}
  &  \mathbb{E}_{\mu} \Big[\mathbb{P}_{ P^{(\mathbb{B}_{n})}}(\hat M \neq M)\Big] \notag \\
  &= \mathbb{E}_{\mu} \Big[\mathbb{P}_{ P^{(\mathbb{B}_{n})}}\big(\hat M \neq M\big |  (I,J,M)=(1,1,1)\big)\Big] \notag \\
    & \leq \mathbb{E}_{\mu}\Big[ \mathbb{P}_{ P^{(\mathbb{B}_{n})}} \big(\mathcal{E}_1(1,1,1) \cup \mathcal{E}_2(1,1,1) \cup \mathcal{E}_3(1,1,1) \big | \notag \\& \qquad \qquad  (I,J,M)=(1,1,1) \big)\Big]. \label{applyunbndiid} &&
\end{flalign}
To upper bound the the RHS of  \eqref{applyunbndiid}, we use the following lemma whose proof is given in Appendix \ref{expdecerrprobproof}.
\begin{lemma} \label{lem:expuppbnd}
If $(R,R_1,R_2)\in\RR_{\geq 0}^3$ satisfy
\begin{align}
R_2+R &< I_P(V;Y|U) \label{eq:errpcond1iid}\\
   R_1+R_2+R &< I_P(U,V;Y), \label{eq:errpcond2iid}
\end{align} 
then there exists a $\zeta(\delta)>0$ such that
\begin{align}
&\mathbb{E}_{\mu}\Big[ \mathbb{P}_{ P^{(\mathbb{B}_{n})}} \big(\mathcal{E}_1(1,1,1) \cup \mathcal{E}_2(1,1,1) \cup \mathcal{E}_3(1,1,1) \big |\notag \\
&\qquad \qquad (I,J,M)=(1,1,1) \big)\Big]\leq e^{-n\zeta(\delta)}. \label{expbndappend}
\end{align}

\end{lemma}
Thus, from \eqref{applyunbndiid} and \eqref{expbndappend}, it follows that 
\begin{align}
    \mathbb{E}_{\mu}\Big[ \mathbb{P}_{P^{(\mathbb{B}_{n})}}\big(\hat M \neq M\big)\Big] \leq e^{-n\zeta(\delta)}  
    \xrightarrow[\ n\ ]{} 0, \label{avgPErandcode}
\end{align}
provided \eqref{eq:errpcond1iid} and \eqref{eq:errpcond2iid} holds.

\textbf{Security analysis:} First consider 
\begin{flalign}
   & I_{P^{(\mathcal{B}_{n})}}(M;\mathbf{Z}) \notag \\
   &\leq I_{P^{(\mathcal{B}_{n})}}(M;I,\mathbf{U},\mathbf{Z}) \notag\\
    & =\kl {P^{(\mathcal{B}_{n})}_{M,I,\mathbf{U},\mathbf{Z}}}{P^{(\mathcal{B}_{n})}_{M}P^{(\mathcal{B}_{n})}_{I,\mathbf{U},\mathbf{Z}}}\notag \\
     & =\kl{P^{(\mathcal{B}_{n})}_{M,I,\mathbf{U},\mathbf{Z}}}{P^{(\mathcal{B}_{n})}_{M,I,\mathbf{U}}P^{(\mathcal{B}_{n})}_{\mathbf{Z}|I,\mathbf{U}}} \notag\\
       &\stackrel{(a)}{=} \kl {P^{(\mathcal{B}_{n})}_{M,I,\mathbf{U}}P^{(\mathcal{B}_{n})}_{\mathbf{Z}|M,I,\mathbf{U}}}{P^{(\mathcal{B}_{n})}_{M,I,\mathbf{U}} P^{\otimes n}_{Z|U}}\notag \\
       & \qquad \qquad -\kl{P^{(\mathcal{B}_{n})}_{I,\mathbf{U},\mathbf{Z}}}{P^{(\mathcal{B}_{n})}_{I,\mathbf{U}} P^{\otimes n}_{Z|U}} \notag\\
      & \stackrel{(b)}{\leq}  \kl{P^{(\mathcal{B}_{n})}_{M,I,\mathbf{U}}P^{(\mathcal{B}_{n})}_{\mathbf{Z}|M,I,\mathbf{U}}}{P^{(\mathcal{B}_{n})}_{M,I,\mathbf{U}} P^{\otimes n}_{Z|U}}\label{MIbndKLiid}\\
      & \stackrel{(c)}{\leq}\mspace{-3 mu}   \max_{m \in \mathcal{M}_n} \mspace{-5 mu} \kl{P^{(\mathcal{B}_{n})}_{I,\mathbf{U}}\mspace{-3 mu}P^{(\mathcal{B}_{n})}_{\mathbf{Z}|M,I,\mathbf{U}}(\cdot|m,\cdot,\cdot)}{P^{(\mathcal{B}_{n})}_{I,\mathbf{U}}\mspace{-3 mu}P^{\otimes n}_{Z|U}}, \label{indepmesind}
    \end{flalign}
    where, 
  \begin{enumerate}[label = (\alph*),leftmargin=5.5mm]
            \item and (c) is since $M \independent (I, \mathbf{U})$;
        \item uses the non-negativity of KL divergence.
    \end{enumerate}
  Maximizing w.r.t. $P_M \in \mathcal{P}(\mathcal{M}_n)$ on both sides of \eqref{indepmesind}, we obtain that
    \begin{flalign}
  & \max_{P_M \in \mathcal{P}(\mathcal{M}_n)}   I_{P^{(\mathcal{B}_{n})}}(M;\mathbf{Z})\notag \\  
   & \leq \mspace{-3 mu}   \max_{m \in \mathcal{M}_n} \mspace{-5 mu} \kl{P^{(\mathcal{B}_{n})}_{I,\mathbf{U}}\mspace{-3 mu}P^{(\mathcal{B}_{n})}_{\mathbf{Z}|M,I,\mathbf{U}}(\cdot|m,\cdot,\cdot)}{P^{(\mathcal{B}_{n})}_{I,\mathbf{U}}\mspace{-3 mu}P^{\otimes n}_{Z|U}}. \label{maxmutinfmsg}
    \end{flalign}
 Note that $P^{(\mathcal{B}_{n})}_{I,\mathbf{U}}P^{(\mathcal{B}_{n})}_{\mathbf{Z}|M,I,\mathbf{U}}(\cdot|m,\cdot,\cdot)\ll P^{(\mathcal{B}_{n})}_{I,\mathbf{U}} P^{\otimes n}_{Z|U}$, where $\ll$ denotes absolute continuity of measures. 
For  PMFs $P,Q$ with a finite support such that $P\ll Q$, $\kl{P}{Q}$ can be upper bounded by the total variation distance $\tv{P}{Q}$\cite[Equation 30]{Cuff-2013}. Defining
\begin{align}
  \theta(m,\mathcal{B}_{n}):=\tv{P^{(\mathcal{B}_{n})}_{I,\mathbf{U}}P^{(\mathcal{B}_{n})}_{\mathbf{Z}|M,I,\mathbf{U}}(\cdot|m,\cdot,\cdot)}{P^{(\mathcal{B}_{n})}_{I,\mathbf{U}} P^{\otimes n}_{Z|U}},  \notag
\end{align}
and applying \cite[Lemma 9]{ZG-arxivv2}, we obtain
 \begin{align}
 & \max_{m \in \mathcal{M}_n} \kl{P^{(\mathcal{B}_{n})}_{I,\mathbf{U}}P^{(\mathcal{B}_{n})}_{\mathbf{Z}|M,I,\mathbf{U}}(\cdot|m,\cdot,\cdot)}{P^{(\mathcal{B}_{n})}_{I,\mathbf{U}} P^{\otimes n}_{Z|U}} \notag \\
  &\leq  \max_{m \in \mathcal{M}_n}\theta(m,\mathcal{B}_{n}) \Big(n \log|\Z|-\log \theta(m,\mathcal{B}_{n})\notag \\
  & \qquad \qquad \qquad +n \log P_{Z|U}^{(\mathsf{min})} \Big),  \label{KLbndTViid}
 \end{align}
 where 
 \begin{align}
 \mspace{-5 mu}P_{Z|U}^{(\mathsf{min})}\mspace{-5 mu}:=\min \{P_{Z|U}(z|u),\mspace{-4 mu}~(z,u) \mspace{-3 mu}\in \mspace{-3 mu}\Z \times \Ucal:P_{Z|U}(z|u)\mspace{-3 mu}>\mspace{-3 mu}0\}. \notag   
 \end{align}
  Thus, showing that there exist $\mathcal{B}_n \in \mathfrak{B}_n$ and $\gamma_1>0$ such that  $\max_{m \in \mathcal{M}_n}\theta(m,\mathcal{B}_{n}) \leq e^{-n \gamma_1}$ for large enough $n$, is sufficient (by \eqref{maxmutinfmsg} and \eqref{KLbndTViid}) to get 
 \begin{align}
 \max_{P_M \in \mathcal{P}(\mathcal{M}_n)}   I_{P^{(\mathcal{B}_{n})}}(M;\mathbf{Z})   \xrightarrow[\ n\ ]{}0. \label{mutinfsecbnd}  
 \end{align}
The existence of such a $\mathcal{B}_n$ is implied by  the following lemma. The lemma restates the outcome of the secrecy analysis from \cite{Goldfeld_SDWTC_journal2020}, providing a double-exponential bound on the probability of an exponentially small deviation of $\max_{m \in \mathcal{M}_n} \theta(m,\mathbb{B}_n)$ from zero. 
\begin{lemma}[Lemma 4 from \cite{Goldfeld_SDWTC_journal2020}] \label{Lem:SSCL}
  If 
  \begin{align}
     R_2>I_P(V;Z|U), \label{eq:SScond} 
  \end{align}
   then there exists $\gamma_1, \gamma_2>0$ such that for all sufficiently large $n$,
  \begin{align}
      \mathbb{P}_{\mu}\left(\max_{m \in \mathcal{M}_n} \theta(m,\mathbb{B}_n) >e^{-n\gamma_1} \right) \leq e^{-e^{n\gamma_2}}. \label{strngsuplemma}
  \end{align}
\end{lemma}
Lemma \ref{Lem:SSCL} follows from the proof of Lemma 4 in \cite{Goldfeld_SDWTC_journal2020}, which is a stronger version of the superposition soft-covering lemma \cite{Cuff-2013}.  The double-exponential bound in \eqref{strngsuplemma} is an implication of Chernoff bound applied to the collection of an exponential number of i.i.d. codewords in the random superposition codebook.   

\textbf{Summary of random coding argument:} Combining \eqref{costavg}, \eqref{avgPErandcode}  and  \eqref{strngsuplemma}, we have shown that 
\begin{subequations}
 \begin{align}
   & \mathbb{E}_{\mu} \Big[ \mathbb{E}_{P_{\mathbf{X}}^{(\mathbb{B}_{n})}}\big[\mathsf{C}_n(\mathbf{X})\big]\Big]\leq b':= b-\gamma' \label{avgcostfin},\\
   & \mathbb{E}_{\mu}\Big[ \mathbb{P}_{P^{(\mathbb{B}_{n})}}(\hat M \neq M)\Big] \leq e^{-n\zeta(\delta)},\\
     &\mathbb{E}_{\mu} \Big[\ind_{\{ \max_{m \in \mathcal{M}_n} \theta(m,\mathbb{B}_n) >e^{-n\gamma_1} \}}\Big] \leq e^{-e^{n\gamma_2}}, 
 \end{align}
 \end{subequations}
 provided \eqref{eq:errpcond1iid}, \eqref{eq:errpcond2iid} and \eqref{eq:SScond} hold.

We next perform a sequence of expurgation steps: first, w.r.t. codebooks and then w.r.t. messages. At the end of this process, we deduce the existence of a single codebook $\mathcal{B}_n$ that satisfies \eqref{permsgcost} and \eqref{relsecconst}. Note that the selection lemma 
of \cite{bloch_barros_2011} or \cite{Goldfeld_SDWTC_journal2020} is not applicable here as the 
RHS of \eqref{avgcostfin} is a constant which does not vanish to zero as required by the lemma. 

\textbf{Expurgation:} For any $\mathfrak{B}'_n \subseteq \mathfrak{B}_n$, let 
\begin{align}
    \bar \mu (\mathfrak{B}'_n):=\sum_{\mathcal{B}_n \in \mathfrak{B}'_n}\mu(\mathcal{B}_n)
\end{align}
be the probability measure on $\mathfrak{B}_n$ induced by $\mu$. Our expurgation technique on the codebooks $\mathcal{B}_n \in \mathfrak{B}_n$ is performed  for each fixed $n$ (sufficiently large) as described in the following steps: \\
\begin{enumerate}[leftmargin=5.5mm]
    \item     \textbf{Codebook expurgation to satisfy average (over messages) cost:} \\[5 pt]
    Expurgate codebooks $\mathcal{B}_n \in \mathfrak{B}_n$ with the highest cost   $\mathbb{E}_{P_{\mathbf{X}}^{(\mathcal{B}_{n})}}\big[\mathsf{C}_n(\mathbf{X})\big]$  to obtain a set $\mathfrak{B}'_n \subset \mathfrak{B}_n$ such that 
    $\frac{1}{n+2} \leq \bar \mu(\mathfrak{B}'_n) <\frac{1}{n+1}$. This is possible for large $n$ since each codebook $\mathcal{B}_n$ has exponentially small probability. We now show that all the codebooks $\mathcal{B}_n \in \mathfrak{B}'_n$ satisfy $\mathbb{E}_{P_{\mathbf{X}}^{(\mathcal{B}_{n})}}\big[\mathsf{C}_n(\mathbf{X})\big]\leq \frac{n+1}{n}b'$. Assume otherwise that there exists $\mathcal{B}_n^* \in \mathfrak{B}'_n$ such that $\mathbb{E}_{P_{\mathbf{X}}^{(\mathcal{B}^*_{n})}}\big[\mathsf{C}_n(\mathbf{X})\big] > \frac{n+1}{n}b'$. Then, we can write
    \begin{flalign}
      \mathbb{E}_{\mu} \Big[ \mathbb{E}_{P_{\mathbf{X}}^{(\mathbb{B}_{n})}}\big[\mathsf{C}_n(\mathbf{X})\big] \Big] &=\sum_{ \mathcal{B}_n \in \mathfrak{B}_n } \mu(\mathcal{B}_n)\mathbb{E}_{P_{\mathbf{X}}^{(\mathcal{B}_{n})}}\big[\mathsf{C}_n(\mathbf{X})\big] \notag \\
      &\geq  \sum_{ \mathcal{B}_n \in \mathfrak{B}_n \setminus \mathfrak{B}'_n } \mu(\mathcal{B}_n)\mathbb{E}_{P_{\mathbf{X}}^{(\mathcal{B}_{n})}}\big[\mathsf{C}_n(\mathbf{X})\big] \notag \\
           &\stackrel{(a)}{\geq} \sum_{ \mathcal{B}_n \in \mathfrak{B}_n \setminus \mathfrak{B}'_n } \mu(\mathcal{B}_n)\mathbb{E}_{P_{\mathbf{X}}^{(\mathcal{B}^*_{n})}}\big[\mathsf{C}_n(\mathbf{X})\big] \notag \\
      &>  \sum_{ \mathcal{B}_n \in \mathfrak{B}_n \setminus \mathfrak{B}'_n } \mu(\mathcal{B}_n) \frac{n+1}{n}b' \notag\\
      & \stackrel{(b)}{\geq} \frac{n}{n+1} \frac{n+1}{n}b'=b', \label{Eq:exp-cont} &&
    \end{flalign}
    where 
 \begin{enumerate}[label = (\alph*),leftmargin=5.5mm]
        \item is because by the expurgation procedure,   $\mathbb{E}_{P_{\mathbf{X}}^{(\mathcal{B}_{n})}}\big[\mathsf{C}_n(\mathbf{X})\big] \geq \mathbb{E}_{P_{\mathbf{X}}^{(\mathcal{B}^*_{n})}}\big[\mathsf{C}_n(\mathbf{X})\big]$ for every $ \mathcal{B}_n \in \mathfrak{B}_n \setminus \mathfrak{B}'_n$  ;
        \item is since $\bar \mu( \mathfrak{B}_n \setminus \mathfrak{B}'_n)  \geq \frac{n}{n+1}$.
    \end{enumerate}
    Eqn. \eqref{Eq:exp-cont} contradicts \eqref{avgcostfin}, and hence  $\mathbb{E}_{P_{\mathbf{X}}^{(\mathcal{B}_{n})}}\big[\mathsf{C}_n(\mathbf{X})\big] \leq  \frac{n+1}{n}b'$ should hold for all $ \mathcal{B}_n \in\mathfrak{B}'_n$.

     Define a PMF $\mu_1\in\cP(\mathfrak{B}_n)$ and its induced probability measure $\bar \mu_1$ on $\mathfrak{B}_n$ by 
    \begin{align} \notag
        \mu_1(\mathcal{B}_n):=\begin{cases}
        \frac{\mu(\mathcal{B}_n)}{ \bar \mu(\mathfrak{B}'_n)}, &\mbox{ if }\mathcal{B}_n \in \mathfrak{B}'_n,\\
        0, & \mbox{ otherwise},
        \end{cases}
    \end{align}
    and 
    \begin{align}
        \bar \mu_1(\tilde{\mathfrak{B}}_n):=\sum_{\mathcal{B}_n \in \tilde{\mathfrak{B}}_n}\mu_1(\mathcal{B}_n),~\tilde{\mathfrak{B}}_n \subseteq  \mathfrak{B}_n,\notag
    \end{align}
respectively. Then, we have 
 \begin{align}
   &  \mathbb{E}_{P_{\mathbf{X}}^{(\mathcal{B}_{n})}}\big[\mathsf{C}_n(\mathbf{X})\big] \leq \frac{n+1}{n}b', ~\forall~\mathcal{B}_n \in \mathfrak{B}'_n, \notag \\
   & \mathbb{E}_{\mu_1} \big[\mathbb{P}_{P^{(\mathbb{B}_{n})}}(\hat M \neq M)\big] \leq \frac{1}{\bar \mu(\mathfrak{B}'_n)}\mathbb{E}_{\mu} \big[\mathbb{P}_{P^{(\mathbb{B}_{n})}}(\hat M \neq M)\big] \notag \\
    &  \leq \frac{e^{-n\zeta(\delta)}}{\bar \mu(\mathfrak{B}'_n)} \leq (n+2)e^{-n\zeta(\delta)},\notag  \\
    & \mathbb{E}_{\mu_1} \Big[\ind_{\{\max_{m \in \mathcal{M}_n} \theta(m,\mathbb{B}_n) >e^{-n\gamma_1} \}}\Big] \notag \\
    &\leq  \frac{1}{\bar \mu(\mathfrak{B}'_n)} \mathbb{E}_{\mu} \Big[\ind_{\{\max_{m \in \mathcal{M}_n} \theta(m,\mathbb{B}_n) >e^{-n\gamma_1} \}}\Big] \notag \\
     &\leq \frac{e^{-e^{n\gamma_2}}}{\bar \mu(\mathfrak{B}'_n)} \leq (n+2)e^{-e^{n\gamma_2}},\notag 
 \end{align}
    \item     \textbf{Codebook expurgation to satisfy average cost, average error probability and semantic-security:}\\[5 pt]
 Expurgate codebooks $\mathcal{B}_n \in \mathfrak{B}'_n$ with the highest average error probability to obtain a set $\mathfrak{B}''_n \subset \mathfrak{B}'_n$ such that $\frac{1}{3} \leq \bar \mu_1(\mathfrak{B}''_n) < \frac{1}{2} $. Then, it follows similarly to step 1 that
 \begin{align}
   \mathbb{P}_{P^{(\mathcal{B}_{n})}}(\hat M \neq M) \leq 2(n+2)e^{-n\zeta(\delta)}, ~\forall~\mathcal{B}_{n} \in \mathfrak{B}''_n. \notag  
 \end{align}
  Define another PMF $\mu_2\in\cP(\mathfrak{B}_n)$ by
     \begin{align} \notag
        \mu_2(\mathcal{B}_n):=\begin{cases}
        \frac{\mu_1(\mathcal{B}_n)}{\bar \mu_1(\mathfrak{B}''_n)}, &\mbox{ if }\mathcal{B}_n \in \mathfrak{B}''_n,\\
        0, & \mbox{ otherwise}.
        \end{cases}
    \end{align}
    Then, we have
 \begin{align}
   &  \mathbb{E}_{P_{\mathbf{X}}^{(\mathcal{B}_{n})}}\big[\mathsf{C}_n(\mathbf{X})\big]\leq \frac{n+1}{n}b', ~\forall~\mathcal{B}_n \in \mathfrak{B}''_n, \notag \\
   &  \mathbb{P}_{P^{(\mathcal{B}_{n})}}(\hat M \neq M)  \leq 2(n+2)e^{-n\zeta(\delta)}, ~\forall~\mathcal{B}_n \in \mathfrak{B}''_n, \notag \\
     &\mathbb{E}_{\mu_2} \Big[\ind_{\{\max_{m \in \mathcal{M}_n} \theta(m,\mathbb{B}_n) >e^{-n\gamma_1} \}}\Big]  \leq 3(n+2)e^{-e^{n\gamma_2}}.\notag 
 \end{align}   
 Perform one more  expurgation step similar to the previous step to obtain a non-empty set of codebooks  $\mathfrak{B}'''_n$ such that for each codebook $\mathcal{B}_n \in \mathfrak{B}'''_n$ and sufficiently large $n$,
  \begin{align}
   &  \mathbb{E}_{P_{\mathbf{X}}^{(\mathcal{B}_{n})}}\big[\mathsf{C}_n(\mathbf{X})\big] \leq \frac{n+1}{n}b', ~\forall~\mathcal{B}_n \in \mathfrak{B}'''_n, \label{avgcostfin2}\\
   &  \mathbb{P}_{P^{(\mathcal{B}_{n})}}(\hat M \neq M)  \leq 2(n+2)e^{-n\zeta(\delta)}, ~\forall~\mathcal{B}_n \mspace{-4 mu} \in \mathfrak{B}'''_n, \label{avgerrpfin2}\\
     &\max_{m \in \mathcal{M}_n} \theta(m,\mathcal{B}_n) \leq e^{-n\gamma_1},~\forall~\mathcal{B}_n \in \mathfrak{B}'''_n. \label{SSfin2}
 \end{align}
 \item \textbf{Message expurgation to satisfy per-message cost, maximal error probability and semantic-security:}\\[5 pt]
 Now, fixing a codebook $\mathcal{B}_{n} \in \mathfrak{B}_n'''$, perform expurgation on the set of messages $\mathcal{M}_n$ to obtain upper bounds on the  per-message cost and maximal error probability, in place of the average (over messages) cost and average (over messages) error probability given in \eqref{avgcostfin2} and \eqref{avgerrpfin2}, respectively. Note that \eqref{avgcostfin2} and \eqref{avgerrpfin2} hold for any message distribution $P_M$. Consider $P_M=\bar P_M$. Let $\alpha \in [\frac{1}{n+2},\frac{1}{n+1})$. Similar to step 1, by  expurgating a $(1-\alpha)$ fraction  of the messages $m \in \mathcal{M}_n$ with the highest cost $\mathbb{E}_{P_{\mathbf{X}|M}^{(\mathcal{B}_{n})}(\cdot|m)}\big[\mathsf{C}_n(\mathbf{X})\big]$ to obtain a set $\mathcal{M}_n' \subset \mathcal{M}_n$, and defining for all $(m,i,j,\mathbf{u},\mathbf{v},\mathbf{x},\mathbf{y},\mathbf{z}, \hat m) \in \mathcal{M}_n \times \mathcal{I}_n \times \mathcal{J}_n \times \Ucal^n \times \V^n \times \X^n \times \Y^n \times \Z^n \times  \hat{\mathcal{M}}_n$, a PMF $\tilde P^{(\mathcal{B}_{n})}$ given by
 \begin{align}
   &  \tilde P^{(\mathcal{B}_{n})}_{M,I,J,\mathbf{U},\mathbf{V},\mathbf{X},\mathbf{Y},\mathbf{Z},\hat M}(m,i,j,\mathbf{u},\mathbf{v},\mathbf{x},\mathbf{y},\mathbf{z}, \hat m)\notag \\
   &:=\begin{cases} 
    \frac{|\mathcal{M}_n|}{|\mathcal{M}'_n|} P^{(\mathcal{B}_{n})}_{M,I,J,\mathbf{U},\mathbf{V},\mathbf{X},\mathbf{Y},\mathbf{Z},\hat M}(m,&\mspace{-15 mu}i,j,\mathbf{u},\mathbf{v},\mathbf{x},\mathbf{y},\mathbf{z}, \hat m),  \\ &\mbox{if } m \in \mathcal{M}_n',\\
     0, & \mbox{otherwise},
     \end{cases} \label{truncdistfin}
 \end{align}
 it follows that 
 \begin{align}
   & \mathbb{E}_{\tilde P_{\mathbf{X}|M}^{(\mathcal{B}_{n})}(\cdot|m)}\big[\mathsf{C}_n(\mathbf{X})\big] \leq  \frac{(n+1)^2}{n^2}b', ~\forall~m \in \mathcal{M}'_n, \notag\\
    &  \mathbb{P}_{\tilde P^{(\mathcal{B}_{n})}}(\hat M \neq M)  \leq 2(n+2)^2e^{-n\zeta(\delta)},\notag \\
     &\max_{m \in \mathcal{M}'_n} \theta(m,\mathcal{B}_n) \leq e^{-n\gamma_1}.\notag
 \end{align}
 Finally, for $\beta \in [\frac{1}{3},\frac 12)$, expurgating a  $1-\beta$ fraction of the messages $m \in \mathcal{M}'_n$ with the highest error probability $\mathbb{P}_{\tilde P^{(\mathcal{B}_{n})}}(\hat M \neq m|M=m)$ and denoting the resulting set by $\mathcal{M}''_n ~(\mathcal{M}''_n \subset \mathcal{M}'_n)$, it follows similar to step 3 that 
  \begin{flalign}
   & \mathbb{E}_{\tilde P_{\mathbf{X}|M}^{(\mathcal{B}_{n})}(\cdot|m)}\big[\mathsf{C}_n(\mathbf{X})\big] \leq  \frac{(n+1)^2}{n^2}b', ~\forall~m \in  \mathcal{M}''_n,\label{finconstexp1}\\
    &  \mathbb{P}_{\tilde P_{\hat M|M}^{(\mathcal{B}_{n})}(\cdot|m)}\mspace{-4 mu}(\hat M \neq m)\mspace{-4 mu}  \leq  4(n+2)^2e^{-n\zeta(\delta)}, ~\mspace{-4 mu}\forall\mspace{-4 mu}~m \in \mathcal{M}''_n,\label{finconstexp2}\\
     &\max_{m \in \mathcal{M}''_n} \theta(m,\mathcal{B}_n) \leq e^{-n\gamma_1}.\label{finconstexp3}
 \end{flalign}
 \textbf{Summary of expurgation steps:}\\[5 pt]
Note that $|\mathcal{M}_n''| = \beta \alpha |\mathcal{M}_n| \geq \frac{e^{nR}}{3(n+2)}$. 
 Thus,  for sufficiently large $n$, we have shown the existence of a codebook $\mathcal{B}_n$ and a $(n,R-\frac 1n \log (3n+6))$ code $c_n(\mathcal{B}_n)=(f_n,g_n)$ with message set $\mathcal{M}''_n$, such that
 \begin{flalign}
     & \mathbb{E}\big[\mathsf{C}_n(\mathbf{X}(m)\big]\notag \\
     &\leq \frac{(n+1)^2}{n^2}b'\notag \\
      &=b-\gamma'+\frac{2n+1}{n^2}\left(b-\gamma'\right)<b,~ \forall ~m \in \mathcal{M}_n'', &&
 \end{flalign}
 and 
 \begin{align}
  & \max \left\lbrace  \max_{m \in \mathcal{M}_n''}e_m\left(c_n(\mathcal{B}_n)\right), \max_{P_M \in \mathcal{P}(\mathcal{M}_n'')}\ell\left(P_M,c_n(\mathcal{B}_n)\right) \right\rbrace \notag \\
  & \leq \epsilon,\label{semsecproberror}
 \end{align}
 provided \eqref{eq:errpcond1iid}, \eqref{eq:errpcond2iid} and \eqref{eq:SScond} holds.
 
 Eliminating $R_1$ and $R_2$ from \eqref{eq:errpcond1iid}, \eqref{eq:errpcond2iid} and \eqref{eq:SScond} via the Fourier-Motzkin elimination   yields  $R <I_P(V;Y|U)-I_P(V;Z|U)$. The proof of \eqref{EQ:main} is completed by noting that $C_\mathsf{sem}(b)$ is a closed set by definition, and $I_P(V;Y|U)-I_P(V;Z|U)$ is a continuous function of $P$.

 \end{enumerate}
\begin{remark}[\textbf{Invariance of secrecy-capacity}]
Theorem \ref{thm:main} states the invariance of CC WTC secrecy-capacity to the employed secrecy metric. The converse proof further shows that capacity remains unchanged if the maximal error probability and/or per message cost constraints are relaxed to average (over messages) constraints of the form $\mathbb{E}_{\bar P_M}\big[e_M(c_n)\big] \leq \epsilon$ and $\mathbb{E}_{\bar P_M}\big[\mathsf{C}_n(\mathbf{X}(M))\big]\leq b$, respectively.
\end{remark}
\begin{remark}[\textbf{Selection lemma}]\label{expurg-rem}
We note that the conclusion of Step 2 of the expurgation procedure can be deduced via the Selection Lemma of \cite{bloch_barros_2011}. However, Steps 1 and 3 seem to require the expurgation argument.
\end{remark}
\subsection{Proof of Proposition \ref{Prop:example}}\label{Sec:Proof:Example}
Since the WTC transition kernel is $P_{Y,Z|\tilde X \tilde Z}=P_{Y|\tilde X,\tilde Z}\ind_{\{Z=\tilde Z\}}$,  $Z=\tilde Z $ with probability one. We henceforth identify $Z$ with $\tilde Z$. We start by showing that
\begin{align}
   \max_{P_{U,V,X} \in \mathcal{H}(0.5) } I_P(V;Y|U)-I_P(V;Z|U) \geq 0.5.  \label{twoauxLB}
\end{align}
Set $U=\tilde  Z \sim \textsf{Ber}(0.5)$, $\tilde X \sim \textsf{Ber}(0.5)$, $\tilde X \independent \tilde Z$,  and $V=X=(\tilde X, \tilde Z)$. This choice satisfies $P_{U,V,X} \in \mathcal{H}(0.5)$, since $U-V-X=(\tilde X,\tilde Z)-(Y,\tilde Z)$ and $\mathbb{E}_P\big[\mathsf{C}(X)\big]:=\mathbb{E}_P\big[\tilde Z\big] = 0.5$. Moreover,
\begin{flalign}
    I_P(V;Y|U)-I_P(V;\tilde Z|U)&=I_P(X;Y|\tilde Z)-I_P(X;\tilde Z| \tilde Z) \notag \\
    &=I_P(X;Y|\tilde Z) \notag \\
    &=0.5, \label{eq:mutinfval} &&
\end{flalign}
where \eqref{eq:mutinfval} is because $I_P(X;Y|\tilde Z)=P_{\tilde Z}(1) H(\tilde X)=0.5$. Hence, \eqref{twoauxLB} holds.

Next, we establish that
\begin{align}
\max_{\substack{P_{V,X}:\\\mathbb{E}_P[\mathsf{C}(X)] \leq 0.5}} I_P(V;Y)-I_P(V;\tilde Z) <0.5.  \label{strictineqsingaux}
\end{align}
For any $P_{V,X}=P_{V, \tilde X, \tilde Z}$ such that $\mathbb{E}_P\big[\tilde Z\big] \leq 0.5$, we have the following chain of inequalities\footnote{We omit the subscript $P$ in the subsequent mutual information and entropy terms as the PMF is $P=P_{V,X}P_{Y,\tilde Z|X}$ 
throughout.}:
\begin{flalign}
 &   I(V;Y)-I(V;\tilde Z) \notag \\
    &=I(\tilde X,\tilde Z,V;Y)-I(\tilde X, \tilde Z;Y|V)-I(V;\tilde Z) \notag \\
     &\stackrel{(a)}{=}I(\tilde X,\tilde Z;Y)-I(\tilde X, \tilde Z;Y|V)-I(V; \tilde Z) \notag \\
     &= I(\tilde X;Y|\tilde Z)-I(\tilde X;Y|\tilde Z,V)-I(V;\tilde Z|Y) \notag \\
     &\stackrel{(b)}{=} P_{\tilde Z}(0)I(\tilde X;N|\tilde Z=0)+P_{\tilde Z}(1)I(Y;Y|\tilde Z=1)\notag \\
     & \qquad -I(\tilde X;Y|\tilde Z,V)-I(V;\tilde Z|Y) \notag  \\
     &\stackrel{(c)}{=}P_{\tilde Z}(1)H(Y|\tilde Z=1)-I(\tilde X;Y|\tilde Z,V)-I(V;\tilde Z|Y)\notag \\
     &\stackrel{(d)}{\leq} 0.5 H(Y|\tilde Z=1)-I(\tilde X;Y|\tilde Z,V)-I(V;\tilde Z|Y)\label{ineqchain1} \\
     & \stackrel{(e)}{\leq} 0.5, \label{mutinfdiffuppbnd} &&
\end{flalign}
where
\begin{enumerate}[label = (\alph*),leftmargin=5.5mm]
    \item is due to the Markov chain $V-(\tilde X,\tilde Z)-(Y,\tilde Z)$;
    \item  follows from the definition of $Y$;
    \item is because $ N \independent (\tilde X,\tilde Z)$;
    \item uses the cost constraint $\mathbb{E}\big[\tilde Z\big] \leq 0.5$;
    \item is by the non-negativity of mutual information and since $Y$ is binary.
\end{enumerate}
Thus,
\begin{align}
\max_{\substack{P_{V,X}:\\\mathbb{E}[\mathsf{C}(X)] \leq 0.5}} I(V;Y)-I(V;\tilde Z)  \leq 0.5. \label{ineqsingauxeq}
\end{align}
 Consequently, \eqref{strictineqsingaux} is violated only if there exist some $X=(\tilde X,\tilde Z)$ and a joint PMF  $P_{V,X,Y,\tilde Z}=P_{V,X}P_{Y,\tilde Z|X}$ such that $\mathbb{E}\big[\tilde Z\big]\leq 0.5$, and the inequalities in \eqref{ineqchain1} and \eqref{mutinfdiffuppbnd} hold with equality, i.e., 
\begin{align}
    I(V;Y)-I(V;\tilde Z)=0.5. \label{eqcontra}
\end{align}
For this to be possible, the following conditions must hold:
\begin{enumerate}
    \item $P_{\tilde Z}(1)=0.5$;
    \item $H(Y|\tilde Z=1)=1$ which means that given $\tilde Z=1$, $Y \sim \textsf{Ber} (0.5)$;
    \item $I(\tilde X;Y|\tilde Z,V)=0$ which implies that $\tilde X-(\tilde Z,V)-Y$ forms a Markov chain;
    \item $I(V;\tilde Z|Y)=0$ which implies that $V-Y-\tilde Z$ forms a Markov chain.
\end{enumerate}

Now, notice that $P_{Y|\tilde Z=0}=P_{Y|\tilde Z=1}=\mathsf{Ber}(0.5)$. Hence, $Y \independent \tilde Z$, which further implies via Condition (4) that $(V,Y) \independent \tilde Z$. Finally, 
\begin{align}
    I(V;Y|\tilde Z=1) \mspace{-2 mu}\geq \mspace{-2 mu} I(\tilde X;Y|\tilde Z=1)\mspace{-2 mu}=\mspace{-2 mu}H(Y|\tilde Z=1)=1. \label{MCcond3}
\end{align}
The inequality in \eqref{MCcond3} is due to Condition (3) above, while the last equality is due to Condition (2). Since $Y$ is binary, $I(V;Y|\tilde Z=1) \leq 1$, and therefore the inequality in \eqref{MCcond3} is an equality.

To conclude, observe that  $(V,Y)\independent \tilde Z$ (shown above) implies that
\begin{align}
    I(V;Y)-I(V;\tilde Z)=I(V;Y|\tilde Z=1)-0=1.
\end{align}
However, $I(V;Y)-I(V;\tilde Z) \leq 0.5$ from \eqref{mutinfdiffuppbnd}. 
This leads to a contradiction, and so \eqref{eqcontra} is invalid. Via \eqref{ineqsingauxeq}, this implies that \eqref{strictineqsingaux} holds. Combining \eqref{strictineqsingaux} with \eqref{twoauxLB} proves Proposition \ref{Prop:example}.

\subsection{Proof of Corollary \ref{Cor:lessnoisy}}\label{Cor:proof}
Fix $P_{U,V,X,Y,Z}=P_{U,V}P_{X|V}P_{Y,Z|X}$, where $Y$ is less noisy than $Z$. We have 
\begin{flalign}
  &  I_P(V;Y|U)-I_P(V;Z|U) \notag\\
    &\stackrel{(a)}{=}I_P(V;Y)-I_P(V;Z)-(I_P(U;Y)-I_P(U;Z)) \notag\\
    &\stackrel{(b)}{=}I_P(X;Y)-I_P(X;Z)-(I_P(X;Y|V)-I_P(X;Z|V))\notag \\
    & \qquad -(I_P(U;Y)-I_P(U;Z)) \notag\\
    & \stackrel{(c)}{\leq} I_P(X;Y)-I_P(X;Z), \notag && 
\end{flalign}
where, $(a)$ and $(b)$ are due to the Markov chain $U-V-X-(Y,Z)$ that holds under PMF $P_{U,V,X,Y,Z}$; and $(c)$ is due to the less noisy assumption which implies that  $I_P(U;Y)-I_P(U;Z) \geq 0$ and $I_P(X;Y|V)-I_P(X;Z|V) \geq 0$ (due to $V-X-(Y,Z)$ under $P_{U,V,X,Y,Z}$). 

Thus, it follows that 
\begin{align}
   \bar C(b) \leq \max_{P_X: \mathbb{E}_P[\mathsf{C}(X)]\leq b} I_P(X;Y)-I_P(X;Z).
\end{align}
The reverse inequality follows trivially by selecting $V=X$ and $U=\emptyset$ in \eqref{SScapmaxexp}, thus proving \eqref{MCseccap}.

If $Z$ is less noisy than $Y$, then $I_P(V;Y|U)-I_P(V;Z|U) \leq 0$ for any  $P_{U,V,X,Y,Z}=P_{U,V}P_{X|V}P_{Y,Z|X}$ which gives $\bar C(b)=0$, due to its non-negativity.
\section{Concluding Remarks}\label{Sec:conclusion}

This paper revisited the classical wiretap channel (WTC) setting with a cost constraint, and showed that achieving its secrecy-capacity generally requires two-layer coding. To do so, we proved optimality of superposition wiretap coding under a cost constraint, and provided a WTC example for which single-layer coding is strictly suboptimal. This stands in contrast to the classic WTC secrecy-capacity result without a cost constraint, that is characterized using a single auxiliary random variable. In many other communication scenarios, imposing a cost constraint on the input amounts to a simple adaptation of the unconstrained case capacity; namely, restricting the optimization to those input distributions that satisfy the constraint in expectation. Our result provides an example where this commonly observed behavior does not hold, and a second auxiliary must be introduced as a result of the added cost constraint. Our main goal was to highlight this important fact and put forth the correct cost constrained (CC) WTC secrecy-capacity characterization, for which non-exact expressions exist in the literature.  

An analogy can be drawn between the secrecy-capacity of a CC WTC and the capacity of channels with action-dependent states \cite{Weissman-2010}. The capacity of the latter with transition kernels $P_{S|A}$ for the action-state pair $(A,S)$ and $P_{Y|X,S}$ for the state-dependent channel is (see \cite[Theorem 1]{Weissman-2010})
\begin{align}
    C_{\mathsf{ADGP}}&=\max I_P(V;Y)-I_P(V;S|A)\notag \\
    &=\max I_P(A;Y)+I_P(V;Y|A)-I_P(V;S|A),\label{adgpcapacity}
\end{align}
where the maximization is taken w.r.t. the joint distribution $P_{A,S,V,X,Y}=P_{A}P_{S|A}P_{V|A,S}P_{X|V,S}P_{Y|X,S}$. Here, $I_P(A;Y)$ can be interpreted as the information gain at the receiver due to the choice of action $A$ (with PMF $P_A$), while the remaining term in \eqref{adgpcapacity} can be thought of as the communication rate achievable over a GP channel $P_{Y|X,S}$ with the action-induced state $S$ distributed as $P_{S}(s)=\sum_{a \in \mathcal{A}} P_{A}(a) P_{S|A}(s|a)$. Here, $A$ takes the role of $U$ in our setting, while the role of $V$ is identical in both settings as the main information carrying auxiliary. However, while $A$ can also be used to convey extra information to the receiver in addition to conditioning the state-dependent channel, $U$ in the CC WTC essentially corresponds to noise added to confuse the eavesdropper, and thus does not carry any information. Accordingly, subtracting  $I_P(A;Y)$ to accommodate this fact leads to the desired analogy between the capacity expressions in the two scenarios.

While this work focused on the discrete setting, an interesting future avenue is to examine the necessity of two auxiliaries for achieving the secrecy-capacity of continuous-alphabet CC WTCs. Since the canonical example of a Gaussian WTC under an average power constraint satisfies the less noisy property (either $Y$ is less noisy than $Z$ or vice-versa), we expect that the capacity expression without any auxiliaries presented in Corollary \ref{Cor:lessnoisy} extends to this case via existing techniques in the literature \cite{CheongHellman-1978,Tyagi-2014,LingLuzzi2014}.  Thus, non-Gaussian channel models, and in particular, ones that cannot be classified as less noisy in general, are the interesting ones to explore. Another possible direction in the continuous channel setup is imposing a peak power constraint on the input. For point-to-point channels, it is known that the peak-constrained capacity achieving distribution is discrete with a finite support \cite{Smith-1971,Shamai-1995,huleihel2018design,Dytso-2020}. Exploring the properties of optimal distributions when an eavesdropper is present seems like a natural extension. It would also be interesting to investigate the above questions in adversarial WTC setting \cite{Boche-2013,WangNaini2016,Goldfeld_AVWTC_semantic_journal2016,Notzel-2016,Wiese-2016,Dey2019,Tahmasbi2020}. A good starting point is to compare the secrecy-capacity with and without cost constraints in special cases where a single-letter solution (without cost) is known \cite{WangNaini2016,Goldfeld_WTCII_semantic_journal2016,Goldfeld_AVWTC_semantic_journal2016,Tahmasbi2020}. Finally, examining the effect of a cost constraint on more complicated networks with secrecy requirements, such as broadcast channels or cloud C-RANs with rate-limited backhaul links  (see, e.g., \cite{Zou-2015} and \cite{Zeyde-2018}) is another potential objective.

\begin{appendices} 
\section{Proof of Lemma \ref{lem:concave}} \label{App1}
 The proof of the cardinality bounds $|\Ucal| \leq |\X|$ and $|\V| \leq |\X|^2$ follows via a standard application of the Eggleston-Fenchel-Carath{\'e}odory Theorem~\cite[Theorem 18]{Eggleston_Convexity1958}, 
 and is omitted.

 Given that $|\cU|$ and $|\cV|$ are finite, the set $\mathcal{H}(b)$ is non-empty (whenever $b \geq \mathsf{c}_{\min}$) and compact. Since $\bar C(b)$ is the supremum of a continuous function $I(V;Y|U)-I(V;Z|U)$ of $P_{U,V,X}$ over $\mathcal{H}(b)$, the supremum is achieved and thus a maximum exists. The fact that $\bar C(b)$ is monotonic and non-decreasing in $b$ follows by its definition.
 
 Finally, to show the concavity of $\bar C(b)$ for $b \geq \mathsf{c}_{\min}$, consider the following. For $i=0,1$, let $b_i \geq \mathsf{c}_{\min}$ and  $P_{U_i,V_i,X_i,Y_i,Z_i}\in \mathcal{P}(\Ucal \times \V \times \X \times \Y \times \Z)$ be a PMF for which
 \begin{align}
 &P_{U_i,V_i,X_i} \in \mathcal{H}(b_i),\\
  & P_{U_i,V_i,X_i,Y_i,Z_i}:=P_{U_i,V_i,X_i}P_{Y_i,Z_i|X_i}:=P_{U_i,V_i,X_i}P_{Y,Z|X}, \notag \\
 &\bar C(b_i):=I_P(V_i;Y_i|U_i)-I_P(V_i;Z_i|U_i).
 \end{align}
Also, let $\tau \in [0,1]$, $Q \sim \mathsf{Ber}(\tau)$ with $\mathcal{Q}=\{0,1\}$, $U_{\tau}:=(U_Q,Q)$, $V_{\tau}:=(V_Q,Q)$ and $X_{\tau}:=X_Q$, and $P_{U_{\tau},V_{\tau},X_{\tau},Y_{\tau},Z_{\tau}} \in \mathcal{P}(\cU_{\tau} \times \cV_{\tau} \times \cX \times \cY \times \cZ)$ be a PMF defined by
\begin{align}
    P_{U_{\tau},V_{\tau},X_{\tau},Y_{\tau},Z_{\tau}}&:=P_{U_{\tau},V_{\tau}}P_{X_{\tau}|V_{\tau}}P_{Y_{\tau},Z_{\tau}|X_{\tau}}\notag \\
    &:=P_{U_{\tau},V_{\tau}}P_{X_{\tau}|V_{\tau}}P_{Y,Z|X}.
\end{align}
Note that $U_{\tau}-V_{\tau}-X_{\tau}-(Y_{\tau},Z_{\tau})$ holds under $P_{U_{\tau},V_{\tau},X_{\tau},Y_{\tau},Z_{\tau}}$, and  $P_{U_{\tau},V_{\tau},X_{\tau}} \in \mathcal{H}((1-\tau) b_0+\tau b_1)$ (by redefining $\mathcal{H}(\cdot)$  in  \eqref{setfeasdist} using $\Ucal\times \mathcal{Q}$ and $\V \times \mathcal{Q}$ in place of $\Ucal$ and $\V$, respectively) since
\begin{align}
  \mathbb{E}_P\left[\mathsf{C}(X_{\tau})\right]\leq (1-\tau) b_0+\tau b_1.  
\end{align}
Then, we have
 \begin{flalign}
 &(1-\tau) \bar C(b_0)+\tau \bar C(b_1) \notag \\
& =(1-\tau) \left(I_P(V_0;Y_0|U_0)-I_P(V_0;Z_0|U_0)
 \right)\notag \\
 & \qquad +\tau \left(I_P(V_1;Y_1|U_1)-I_P(V_1;Z_1|U_1) \right) \notag \\
  &=I_P(V_Q,Q;Y_Q|U_Q,Q)-I_P(V_Q,Q;Z_Q|U_Q,Q)
 \notag \\
 &=I_P(V_{\tau};Y_{\tau}|U_{\tau})-I_P(V_{\tau};Z_{\tau}|U_{\tau})
 \notag \\
    & \leq  \bar C((1-\tau) b_0+\tau b_1),\notag &&
 \end{flalign}
which establishes the concavity of $\bar C(\cdot)$.

\section{Proof of Lemma \ref{lem:expuppbnd}} \label{expdecerrprobproof}
The proof is standard, however, we provide it for completeness. Let  $P^{(\mu)}(\cdot):=\mathbb{E}_\mu\big[P^{(\mathbb{B}_n)}(\cdot)\big]$ and  $\mathbb{P}_{P^{(\mu)}}$ denote the random coding PMF and its induced probability measure, respectively. Also, define
\begin{align}
 \zeta_1^{(n)}(\delta)&:= \inf_{\substack{\nu_{\mathbf{u},\mathbf{v},\mathbf{y}}:\\ (\mathbf{u},\mathbf{v},\mathbf{y}) \notin \typ{P_{U,V,Y}}{\delta}{n}}}\kl{\nu_{\mathbf{u},\mathbf{v},\mathbf{y}} }{P_{U,V,Y}}\notag \\
 & \qquad \qquad \qquad\qquad -|\Ucal||\V||\Y|\frac{\log(n+1)}{n}, \notag
\end{align}
where $\nu_{\mathbf{u},\mathbf{v},\mathbf{y}}$ denotes the empirical PMF of  $(\mathbf{u},\mathbf{v},\mathbf{y})\in\cU^n\times\cV^n\times\cY^n$. Note that for $\delta>0$ and $n$ sufficiently large, $ \zeta_1^{(n)}(\delta)>0$ . 

First, consider the probability of the error event $\mathcal{E}_1(1,1,1)$ averaged over the random codebook $\mathbb{B}_n$. We have
\begin{flalign}
&  \mathbb{E}_{\mu}\Big[\mathbb{P}_{P^{(\mathbb{B}_{n})}}\big(\mathcal{E}_1(1,1,1) \big | (I,J,M)=(1,1,1) \big)\Big]\notag \\
&=\mathbb{P}_{P^{(\mu)}}\Big(\big(\mathbf{U}(1),\mathbf{V}(1,1,1),\mathbf{Y}\big) \notin \typ{P_{U,V,Y}}{\delta}{n}\Big)\notag \\
  &\leq e^{-n  \zeta_1^{(n)}(\delta)} \underset{n}{\longrightarrow}0 , \label{eq:mt1}&&
\end{flalign}
 where the inequality in \eqref{eq:mt1} follows from Lemmas 2.2 and 2.6 in \cite{Csiszar-Korner}. 

Next, we analyze the probability of the error event $\mathcal{E}_2(1,1,1)$ averaged over $\mathbb{B}_n$. Note that for $(j,m) \neq (1,1)$ and  sufficiently large $n$,
\begin{flalign}
& \PP_{P^{(\mu)}}\Big(\big(\mathbf{U}(1),\mathbf{V}(1,j,m),\mathbf{Y}\big) \in \typ{P_{U,V,Y}}{\delta}{n} \big |\notag \\
&\qquad \qquad  (I,J,M)=(1,1,1)\Big)\notag \\
&=\sum_{(\mathbf{u},\mathbf{v},\mathbf{y})~ \in ~\typ{P_{U,V,Y}}{\delta}{n}} P_{U,V}^{\otimes n}(\mathbf{u},\mathbf{v}) P_{Y|U}^{\otimes n}(\mathbf{y}|\mathbf{u})\notag    \\
& \leq \sum_{(\mathbf{u},\mathbf{v},\mathbf{y})~ \in ~\typ{P_{U,V,Y}}{\delta}{n}}e^{-n\left(H_P(U,V)+H_P(Y|U)-O(\delta)\right)}\notag \\
& \leq e^{-n\left(H_P(U,V)+H_P(Y|U)-H_P(U,V,Y)-O(\delta)\right)}\notag \\
&=e^{-n\left(I_P(V;Y|U)-O(\delta)\right)}.\notag &&
\end{flalign}
Hence,
\begin{flalign}
    & \mathbb{E}_{\mu}\Big[ \mathbb{P}_{P^{(\mathbb{B}_{n})}}\big(\mathcal{E}_2(1,1,1) \big | (I,J,M)=(1,1,1) \big)\Big] \notag \\
    &\leq \sum_{\substack{(j,m)  \in~ \mathcal{J}_n \times \mathcal{M}_n:\\(j,m) \neq (1,1)}} \mathbb{P}_{P^{(\mu)}}\Big(\big(\mathbf{U}(1),\mathbf{V}(1,j,m),\mathbf{Y}\big) \in \notag \\ & \qquad \qquad \qquad \qquad \typ{P_{U,V,Y}}{\delta}{n} \big | (I,J,M)=(1,1,1)\Big) \notag \\
    & \leq e^{n(R_2+R-I_P(V;Y|U)+O(\delta))}. \notag &&
\end{flalign}
This implies that for $\delta$ sufficiently small and $n$ large enough, there exists $\zeta_2(\delta)>0$ such that
\begin{flalign}
 & \mathbb{E}_{\mu}\Big[ \mathbb{P}_{P^{(\mathbb{B}_{n})}}\big(\mathcal{E}_2(1,1,1) \big | (I,J,M)=(1,1,1) \big)\Big] \notag \\
 &\leq e^{-n \zeta_2(\delta)}  \underset{n}{\longrightarrow} 0,\label{eq:mt2} &&
\end{flalign}
provided \eqref{eq:errpcond1iid} holds.

Finally, consider the third error event $\mathcal{E}_3(1,1,1)$. We have for $i \neq 1$ and sufficiently large $n$ that
\begin{flalign}
 & \PP_{P^{(\mu)}}\Big(\big(\mathbf{U}(i),\mathbf{V}(i,j,m),\mathbf{Y}\big) \in \typ{P_{U,V,Y}}{\delta}{n} \big |\notag \\
 &\qquad \qquad (I,J,M)=(1,1,1)\Big)\notag \\
&=\sum_{(\mathbf{u},\mathbf{v},\mathbf{y})~ \in ~\typ{P_{U,V,Y}}{\delta}{n}} P_{U,V}^{\otimes n}(\mathbf{u},\mathbf{v}) P_{Y}^{\otimes n}(\mathbf{y})\notag \\
& \leq \sum_{(\mathbf{u},\mathbf{v},\mathbf{y})~ \in ~\typ{P_{U,V,Y}}{\delta}{n}}e^{-n\left(H_P(U,V)+H_P(Y)-O(\delta)\right)}\notag \\
& \leq e^{-n\left(H_P(U,V)+H_P(Y)-H_P(U,V,Y)-O(\delta)\right)}\notag \\
&=e^{-n\left(I_P(U,V;Y)-O(\delta)\right)}.\notag &&   
\end{flalign}
Hence,
\begin{flalign}
    & \mathbb{E}_{\mu}\Big[ \mathbb{P}_{P^{(\mathbb{B}_{n})}}\big(\mathcal{E}_3(1,1,1) \big | (I,J,M)=(1,1,1) \big)\Big] \notag \\
    &\leq \sum_{\substack{(i,j,m)  \in~\mathcal{I}_n \times  \mathcal{J}_n \times \mathcal{M}_n:\\i \neq 1}} \mathbb{P}_{P^{(\mu)}}\Big(\big(\mathbf{U}(i),\mathbf{V}(i,j,m),\mathbf{Y}\big) \in \notag \\ & \qquad \qquad \qquad \qquad \typ{P_{U,V,Y}}{\delta}{n} \big | (I,J,M)=(1,1,1)\Big) \notag \\
    & \leq e^{n(R_1+R_2+R-I_P(U,V;Y)+O(\delta))}. \notag &&
\end{flalign}
Thus, it follows  that for  $\delta$ sufficiently small and $n$ large enough, there exists $\zeta_3(\delta)>0$ such that
\begin{flalign}
     &\mathbb{E}_{\mu}\Big[ \mathbb{P}_{P^{(\mathbb{B}_{n})}}\big(\mathcal{E}_3(1,1,1) \big | (I,J,M)=(1,1,1) \big)\Big] \notag \\
     &\leq e^{-n \zeta_3(\delta)}  \underset{n}{\longrightarrow} 0, \label{eq:mt3}&&
\end{flalign}
provided \eqref{eq:errpcond2iid} holds. The claim in the lemma follows from \eqref{eq:mt1}, \eqref{eq:mt2} and \eqref{eq:mt3}  via the union bound on probability applied to the left hand side of \eqref{expbndappend}.

\end{appendices}
\bibliographystyle{IEEEtran}
\bibliography{ref}

\begin{thebibliography}{10}
\providecommand{\url}[1]{#1}
\csname url@samestyle\endcsname
\providecommand{\newblock}{\relax}
\providecommand{\bibinfo}[2]{#2}
\providecommand{\BIBentrySTDinterwordspacing}{\spaceskip=0pt\relax}
\providecommand{\BIBentryALTinterwordstretchfactor}{4}
\providecommand{\BIBentryALTinterwordspacing}{\spaceskip=\fontdimen2\font plus
\BIBentryALTinterwordstretchfactor\fontdimen3\font minus
  \fontdimen4\font\relax}
\providecommand{\BIBforeignlanguage}[2]{{%
\expandafter\ifx\csname l@#1\endcsname\relax
\typeout{** WARNING: IEEEtran.bst: No hyphenation pattern has been}%
\typeout{** loaded for the language `#1'. Using the pattern for}%
\typeout{** the default language instead.}%
\else
\language=\csname l@#1\endcsname
\fi
#2}}
\providecommand{\BIBdecl}{\relax}
\BIBdecl

\bibitem{Wyner_Wiretap1975}
A.~D. Wyner, ``The wire-tap channel,'' \emph{Bell Sys. Techn.}, vol.~54, no.~8,
  pp. 1355--1387, Oct. 1975.

\bibitem{Csiszar_Korner_BCconfidential1978}
I.~Csisz{\'a}r and J.~K{\"o}rner, ``Broadcast channels with confidential
  messages,'' \emph{IEEE Trans. Inf. Theory}, vol.~24, no.~3, pp. 339--348, May
  1978.

\bibitem{Mukherjee-2014}
A.~{Mukherjee}, S.~A.~A. {Fakoorian}, J.~{Huang}, and A.~L. {Swindlehurst},
  ``Principles of physical layer security in multiuser wireless networks: A
  survey,'' \emph{IEEE Communications Surveys Tutorials}, vol.~16, no.~3, pp.
  1550--1573, Feb. 2014.

\bibitem{Wu-2018}
Y.~{Wu}, A.~{Khisti}, C.~{Xiao}, G.~{Caire}, K.~{Wong}, and X.~{Gao}, ``A
  survey of physical layer security techniques for 5{G} wireless networks and
  challenges ahead,'' \emph{IEEE Journal on Selected Areas in Communications},
  vol.~36, no.~4, pp. 679--695, Apr. 2018.

\bibitem{Hamamreh-2019}
J.~M. {Hamamreh}, H.~M. {Furqan}, and H.~{Arslan}, ``Classifications and
  applications of physical layer security techniques for confidentiality: A
  comprehensive survey,'' \emph{IEEE Communications Surveys Tutorials},
  vol.~21, no.~2, pp. 1773--1828, Oct. 2019.

\bibitem{Liang-book}
Y.~Liang, H.~V. Poor, and S.~Shamai~(Shitz), ``Information theoretic
  security,'' \emph{Found. Trends Commun. Inf. Theory}, vol.~5, no. 4–5, p.
  355–580, Apr. 2009.

\bibitem{BL-2013}
M.~R. Bloch and J.~N. Laneman, ``Strong secrecy from channel resolvability,''
  \emph{{IEEE} Trans. Inf. Theory}, vol.~59, no.~12, pp. 8077--8098, Dec. 2013.

\bibitem{HES-2014}
T.~S. Han, H.~Endo, and M.~Sasaki, ``Reliability and secrecy functions of the
  wiretap channel under cost constraint,'' \emph{{IEEE} Trans. Inf. Theory},
  vol.~60, no.~11, pp. 6819--6843, Nov. 2014.

\bibitem{Goldfeld_WT-GP_analogy_journal2017}
Z.~Goldfeld and H.~H. Permuter, ``Wiretap and {Gelfand-Pinsker} channels
  analogy and its applications,'' \emph{IEEE Trans. Inf. Theory}, vol.~65,
  no.~8, pp. 4979--4996, Aug. 2019.

\bibitem{GelfandPinsker1980}
S.~I. Gelfand and M.~S. Pinsker, ``Coding for channels with random
  parameters,'' \emph{Prob. of Control and Inf. Theory}, vol.~9, no.~1, pp.
  19--31, 1980.

\bibitem{BarronChenWornell2003}
R.~J. {Barron}, B.~Chen, and G.~W. {Wornell}, ``The duality between information
  embedding and source coding with side information and some applications,''
  \emph{{IEEE} Trans. Inf. Theory}, vol.~49, no.~5, pp. 1159--1180, May 2003.

\bibitem{PradhanChouRamachandran2003}
S.~Pradhan, J.~Chou, and K.~Ramchandran, ``Duality between source coding and
  channel coding and its extension to the side information case,'' \emph{{IEEE}
  Trans. Inf. Theory}, vol.~49, no.~5, pp. 1181--1203, May 2003.

\bibitem{Maurer_Strong_Secrecy_Chapter1994}
U.~Maurer, \emph{Communications and Cryptography: Two Sides of One
  Tapestry}.\hskip 1em plus 0.5em minus 0.4em\relax Norwell, MA, USA: Springer
  US, 1994, ch. The Strong Secret Key Rate of Discrete Random Triples, pp.
  271--285.

\bibitem{Csiszar_Strong_Secrecy1996}
I.~Csisz{\'a}r, ``Almost independence and secrecy capacity,'' \emph{Prob. Inf.
  Trans.}, vol.~32, no.~1, pp. 40--47, Jan.-Mar. 1996.

\bibitem{Vardy_Semantic_WTC2012}
M.~Bellare, S.~Tessaro, and A.~Vardy, ``A cryptographic treatment of the
  wiretap channel,'' in \emph{Proc. Adv. Crypto. (CRYPTO 2012)}, Santa Barbara,
  CA, USA, Aug. 2012.

\bibitem{Csiszar-Korner}
I.~Csisz\'{a}r and J.~K\"{o}rner, \emph{Information Theory: Coding Theorems for
  Discrete Memoryless Systems}.\hskip 1em plus 0.5em minus 0.4em\relax
  Cambridge University Press, 2011.

\bibitem{Gallagerbook}
R.~G. Gallager, \emph{Information Theory and Reliable Communication}.\hskip 1em
  plus 0.5em minus 0.4em\relax John Wiley and Sons (New York), 1968.

\bibitem{Elgamalkim}
A.~E. Gamal and Y.~H. Kim, \emph{Network Information theory}.\hskip 1em plus
  0.5em minus 0.4em\relax Cambridge University Press, 2011.

\bibitem{Han-infospectrum}
T.~S. Han, \emph{Information-Spectrum Methods in Information Theory}.\hskip 1em
  plus 0.5em minus 0.4em\relax Springer-Verlag Berlin Heidelberg, 2003.

\bibitem{Verdu-1990}
S.~{Verdu}, ``On channel capacity per unit cost,'' \emph{{IEEE} Trans. Inf.
  Theory}, vol.~36, no.~5, pp. 1019--1030, Sep. 1990.

\bibitem{Prelov-1993}
V.~V. {Prelov} and E.~C. {van der Meulen}, ``An asymptotic expression for the
  information and capacity of a multidimensional channel with weak input
  signals,'' \emph{{IEEE} Trans. Inf. Theory}, vol.~39, no.~5, pp. 1728--1735,
  Sep. 1993.

\bibitem{Hajek-2002}
B.~{Hajek} and V.~G. {Subramanian}, ``Capacity and reliability function for
  small peak signal constraints,'' \emph{{IEEE} Trans. Inf. Theory}, vol.~48,
  no.~4, pp. 828--839, Aug. 2002.

\bibitem{ElGamal-2006}
A.~{El Gamal}, M.~{Mohseni}, and S.~{Zahedi}, ``Bounds on capacity and minimum
  energy-per-bit for {AWGN} relay channels,'' \emph{{IEEE} Trans. Inf. Theory},
  vol.~52, no.~4, pp. 1545--1561, Apr. 2006.

\bibitem{Yagli-2019}
S.~{Yagli} and P.~{Cuff}, ``Exact exponent for soft covering,'' \emph{{IEEE}
  Trans. Inf. Theory}, vol.~65, no.~10, pp. 6234--6262, Oct. 2019.

\bibitem{Korner-Marton-1977}
J.~K\"{o}rner and K.~Marton, \emph{Colloquia Mathematica Societatis J\'{a}nos
  Bolyai}.\hskip 1em plus 0.5em minus 0.4em\relax Keszthely, Hungary:
  North-Holland, Amsterdam, 1977, ch. Comparison of two noisy channels, pp.
  411--423.

\bibitem{CoverThomas}
T.~M. Cover and J.~A. Thomas, \emph{Elements of Information Theory}.\hskip 1em
  plus 0.5em minus 0.4em\relax Wiley (New York), 1991.

\bibitem{Cuff-2013}
P.~Cuff, ``Distributed channel synthesis,'' \emph{{IEEE} Trans. Inf. Theory},
  vol.~59, no.~11, pp. 7071--7096, Nov. 2013.

\bibitem{ZG-arxivv2}
Z.~Goldfeld, P.~Cuff, and H.~H. Permuter, ``Wiretap channels with random states
  non-causally available at the encoder,'' \emph{arXiv preprint
  arXiv:1608.00743v2}, 2018.

\bibitem{Goldfeld_SDWTC_journal2020}
------, ``Wiretap channel with random states non-causally available at the
  encoder,'' \emph{IEEE Trans. Inf. Theory}, vol.~66, no.~3, pp. 1497--1519,
  Mar. 2020.

\bibitem{bloch_barros_2011}
M.~Bloch and J.~Barros, \emph{Physical-Layer Security: From Information Theory
  to Security Engineering}.\hskip 1em plus 0.5em minus 0.4em\relax Cambridge
  University Press, 2011.

\bibitem{Weissman-2010}
T.~{Weissman}, ``Capacity of channels with action-dependent states,''
  \emph{{IEEE} Trans. Inf. Theory}, vol.~56, no.~11, pp. 5396--5411, Oct. 2010.

\bibitem{CheongHellman-1978}
S.~K. Cheong and M.~E. Hellman, ``The {G}aussian wire-tap channel,''
  \emph{{IEEE} Trans. Inf. Theory}, vol.~24, no.~4, pp. 451--456, Jul. 1978.

\bibitem{Tyagi-2014}
H.~{Tyagi} and A.~{Vardy}, ``Explicit capacity-achieving coding scheme for the
  {G}aussian wiretap channel,'' in \emph{IEEE International Symposium on
  Information Theory (ISIT)}, Honolulu, HI, USA, Aug. 2014, pp. 956--960.

\bibitem{LingLuzzi2014}
C.~Ling, L.~Luzzi, J.~C. Belfiore, and D.~Stehl\'{e}, ``Semantically secure
  lattice codes for the {G}aussian wiretap channel,'' \emph{{IEEE} Trans. Inf.
  Theory}, vol.~60, no.~10, pp. 6399--6416, Oct. 2014.

\bibitem{Smith-1971}
J.~G. {Smith}, ``The information capacity of amplitude- and
  variance-constrained scalar {G}aussian channels,'' \emph{Information and
  Control}, vol.~18, no.~3, pp. 203 -- 219, 1971.

\bibitem{Shamai-1995}
S.~{Shamai} and I.~{Bar-David}, ``The capacity of average and
  peak-power-limited quadrature {G}aussian channels,'' \emph{{IEEE} Trans. Inf.
  Theory}, vol.~41, no.~4, pp. 1060--1071, Jul. 1995.

\bibitem{huleihel2018design}
W.~Huleihel, Z.~Goldfeld, T.~Koch, M.~Madiman, and M.~M{\'e}dard, ``Design of
  discrete constellations for peak-power-limited complex {G}aussian channels,''
  in \emph{IEEE International Symposium on Information Theory (ISIT)}, Vail,
  CO, USA, Jun. 2018, pp. 556--560.

\bibitem{Dytso-2020}
A.~{Dytso}, S.~{Yagli}, H.~V. {Poor}, and S.~{Shamai Shitz}, ``The capacity
  achieving distribution for the amplitude constrained additive {G}aussian
  channel: An upper bound on the number of mass points,'' \emph{{IEEE} Trans.
  Inf. Theory}, vol.~66, no.~4, pp. 2006--2022, Oct. 2020.

\bibitem{Boche-2013}
H.~{Boche} and R.~F. {Schaefer}, ``Capacity results and super-activation for
  wiretap channels with active wiretappers,'' \emph{IEEE Trans. Inf. Forensics
  and Security}, vol.~8, no.~9, pp. 1482--1496, Aug. 2013.

\bibitem{WangNaini2016}
P.~{Wang} and R.~S. {Naini}, ``A model for adversarial wiretap channels,''
  \emph{{IEEE} Trans. Inf. Theory}, vol.~62, no.~2, pp. 970--983, Feb. 2016.

\bibitem{Goldfeld_AVWTC_semantic_journal2016}
Z.~Goldfeld, P.~Cuff, and H.~H. Permuter, ``Arbitrarily varying wiretap
  channels with type constrained states,'' \emph{IEEE Trans. Inf. Theory},
  vol.~62, no.~12, pp. 7216--7244, Dec. 2016.

\bibitem{Notzel-2016}
J.~{Nötzel}, M.~{Wiese}, and H.~{Boche}, ``The arbitrarily varying wiretap
  channel—secret randomness, stability, and super-activation,'' \emph{{IEEE}
  Trans. Inf. Theory}, vol.~62, no.~6, pp. 3504--3531, Apr. 2016.

\bibitem{Wiese-2016}
M.~{Wiese}, J.~{Nötzel}, and H.~{Boche}, ``A channel under simultaneous
  jamming and eavesdropping attack—correlated random coding capacities under
  strong secrecy criteria,'' \emph{{IEEE} Trans. Inf. Theory}, vol.~62, no.~7,
  pp. 3844--3862, May 2016.

\bibitem{Dey2019}
B.~K. {Dey}, S.~{Jaggi}, and M.~{Langberg}, ``Sufficiently myopic adversaries
  are blind,'' \emph{{IEEE} Trans. Inf. Theory}, vol.~65, no.~9, pp.
  5718--5736, Sep. 2019.

\bibitem{Tahmasbi2020}
M.~{Tahmasbi}, M.~R. {Bloch}, and A.~{Yener}, ``Learning an adversary’s
  actions for secret communication,'' \emph{{IEEE} Trans. Inf. Theory},
  vol.~66, no.~3, pp. 1607--1624, Mar. 2020.

\bibitem{Goldfeld_WTCII_semantic_journal2016}
Z.~Goldfeld, P.~Cuff, and H.~H. Permuter, ``Semantic-security capacity for
  wiretap channels of type {II},'' \emph{IEEE Trans. Inf. Theory}, vol.~62,
  no.~7, pp. 3863--3879, Jul. 2016.

\bibitem{Zou-2015}
S.~{Zou}, Y.~{Liang}, L.~{Lai}, H.~V. {Poor}, and S.~{Shamai}, ``Broadcast
  networks with layered decoding and layered secrecy: Theory and
  applications,'' \emph{Proceedings of the IEEE}, vol. 103, no.~10, pp.
  1841--1856, Sep. 2015.

\bibitem{Zeyde-2018}
M.~{Zeyde}, O.~{Simeone}, and S.~{Shamai}, ``Confidential communication in
  {C-RAN} systems with infrastructure sharing,'' in \emph{2018 IEEE
  International Conference on the Science of Electrical Engineering in Israel
  (ICSEE)}, Elat, Israel, Dec. 2018, pp. 1--5.

\bibitem{Eggleston_Convexity1958}
H.~G. Eggleston, \emph{Convexity}, 6th~ed.\hskip 1em plus 0.5em minus
  0.4em\relax Cambridge, U.K.: Cambridge Univ. Press, 1958.

\end{thebibliography}

\end{document}